\documentclass{aastex631}

\usepackage{hyperref}
\usepackage{graphicx}
\usepackage{subfigure}
\usepackage[graphicx]{realboxes}
\usepackage{appendix}
\usepackage{booktabs}
\usepackage{tabularx}
\usepackage{multirow}
\usepackage{amsmath}
\usepackage{hhline}
\usepackage[clockwise]{rotating}
\usepackage{multirow}
\usepackage{natbib}
\usepackage{makecell}
\usepackage{float}
\usepackage{subfigure}
\usepackage{chngcntr}

\shorttitle{Kinematics of YSOs Under Various Stellar Feedback}
\shortauthors{Yang et al.}

\graphicspath{{./}{figures/}}
\begin{document}

\title{Kinematics of Young Stellar Objects Under Various Stellar Feedback}

\author{Longhui Yang}
\affiliation{Guangxi Key Laboratory for Relativistic Astrophysics, School of Physical Science and Technology, Guangxi University, Nanning 530004, China}

\author{Dejian Liu}
\affiliation{Purple Mountain Observatory, Chinese Academy of Sciences, Nanjing 210023, People's Republic of China}
\affiliation{College of Science, China Three Gorges University, Yichang, 443000, China}

\author{Chaojie Hao}
\affiliation{Purple Mountain Observatory, Chinese Academy of Sciences, Nanjing 210023, People's Republic of China}

\author{Zehao Lin}
\affiliation{Purple Mountain Observatory, Chinese Academy of Sciences, Nanjing 210023, People's Republic of China}

\author{YingJie Li}
\affiliation{Purple Mountain Observatory, Chinese Academy of Sciences, Nanjing 210023, People's Republic of China}

\author{Yiwei Dong}
\affiliation{Purple Mountain Observatory, Chinese Academy of Sciences, Nanjing 210023, People's Republic of China}
\affiliation{School of Astronomy and Space Science, University of Science and Technology of China, Hefei 230026, People's Republic of China}

\author{Zu-Jia Lu}
\affiliation{Guangxi Key Laboratory for Relativistic Astrophysics, School of Physical Science and Technology, Guangxi University, Nanning 530004, China}

\author{En-Wei Liang}\thanks{lew@gxu.edu.cn (EWL)}
\affiliation{Guangxi Key Laboratory for Relativistic Astrophysics, School of Physical Science and Technology, Guangxi University, Nanning 530004, China}

\author{Y. Xu}\thanks{xuye@pmo.ac.cn}
\affiliation{Purple Mountain Observatory, Chinese Academy of Sciences, Nanjing 210023, People's Republic of China}
\affiliation{School of Astronomy and Space Science, University of Science and Technology of China, Hefei 230026, People's Republic of China}



\begin{abstract}
	
Based on the Gaia Data Release 3 and APOGEE datasets, we investigate the kinematic differences between young stellar objects (YSOs) and their parent clouds in five nearby star-forming regions. Overall, the 1D velocity differences between Class II YSOs and their parent molecular cloud range from [0, 1.4] km s$^{-1}$. In feedback environments dominated by outflows, massive stars, and supernova feedback, the corresponding velocity differences range from [0, 1.4] km s$^{-1}$, [0.1, 0.4] km s$^{-1}$, and [0.1, 1] km s$^{-1}$, respectively. These results indicate that YSO kinematics are not significantly affected by these different types of feedback environment. Additionally, compared to the Class II YSOs, Class III YSOs have slightly larger velocities and dispersions. 

\end{abstract}
\keywords{Stellar kinematics (1608) -- Molecular clouds (1072) -- Young stellar objects (1834)}


\section{introduction} \label{sec:intro}

The vast majority of stars form in clusters within molecular clouds, subsequently evolving into star clusters, associations, or isolated stars, and eventually dissolving into the Galactic gravitational field \citep[e.g.,][]{Zeeuw1999,Lada2003,Buckner2019,Krause2020}. Stars form in molecular clouds through gravitational collapse, and are influenced by turbulence \citep[e.g.,][]{Hennebelle2012,Guerrero-Gamboa2020}, stellar feedback \citep[e.g.,][]{Lada2003,Rey-Raposo2017}, and magnetic fields \citep[e.g.,][]{Ostriker2001,McKee2007}. Currently, our understanding of the kinematic processes of star formation remains quite limited. Young stellar objects (YSOs), which are very young and still associated with their parent molecular clouds, can potentially serve as tracers to reveal the kinematics of the star formation process.

In recent studies, researchers have made significant progress in understanding the relationship between molecular clouds and star formation by combining observations of YSOs and molecular gas. For example, Gaia astrometry data of YSOs have been used to explore the structure of molecular clouds \citep[e.g.,][]{Kounkel2022b,Tu2022}, the history of star formation \citep[e.g.,][]{Gro2021A&A...647A..91G,Krolikowski2021}, and the expansion of young star clusters \citep{Kuhn2019ApJ...870...32K}. In these studies, researchers generally assume that the parallaxes and proper motions of the YSOs can serve as proxies for the distances and proper motions of the molecular gas therein. However, due to stellar feedback, the kinematics of the YSOs may not be entirely consistent with that of their parent molecular clouds.

During the star formation process, various feedback environments such as protostellar outflows \citep[e.g.,][]{Arce2010ApJ...715.1170A,Dionatos2017}, stellar winds from massive stars and H~\textsc{ii} regions \citep[e.g.,][]{Pabst2019Natur.565..618P,Grudic2022}, and supernova explosions \citep[e.g.,][]{Wareing2017,Krumholz2019ARA&A..57..227K} inject momentum into star-forming regions. These processes are theoretically capable of displacing surrounding material and potentially influencing the kinematics of newly formed stars within the cloud. However, to date, there remains a lack of research comparing the kinematic characteristics of YSOs with their parent molecular clouds, which is the purpose of our research. By revealing the kinematic differences and their extent between YSOs and their parent molecular clouds, we aim to provide a foundation for studying star formation and kinematic evolution.

Now, we have the opportunity to explore the kinematic characteristics of YSOs and compare them with the motions of their parent molecular clouds. First, molecular line surveys provide us with distribution and kinematic information of molecular clouds \citep{Dame2001ApJ...547..792D}. Second, a large number of YSOs have been identified through previous surveys, such as \citet{Gutermuth2009,Megeath2012,Furlan2016}. Thus, by crossmatching infrared-selected YSOs with Gaia's high-precision astrometric data \citep{Gaia2023}, we can obtain kinematic information for a significant number of YSOs.

The structure of this paper is as follows: Section \ref{Data} describes the star-forming regions and the corresponding YSO samples. Section \ref{results} presents the distribution and kinematics of the star-forming regions and YSOs. Section \ref{discussion} provides a statistical analysis of the kinematic characteristics of YSOs in different feedback environments. Finally, Section \ref{conclusion} provides a summary.

\section{Data} 
\label{Data}
\subsection{Molecular Clouds}

Considering Gaia's astrometric precision and the number of YSOs in our sample, we first focus on star-forming regions within 500 pc of the Sun, including Orion~A, Orion~B, Perseus, Taurus, and $\lambda$ Orionis. A significant number of YSOs have been identified in these star-forming regions \citep[e.g.,][]{Gutermuth2009,Megeath2012,Furlan2016}. On the other hand, these regions encompass various stellar feedback environments, such as outflows, photoionization, radiation pressure, thermal stellar winds from H~\textsc{ii} regions, and supernova explosions \citep[e.g.,][]{Dolan2002AJ....123..387D,Arce2010ApJ...715.1170A,Pabst2019Natur.565..618P}. This allows us not only to study the kinematic differences between YSOs and their parent molecular clouds but also to reveal the conditions in their environments dominated by different feedback mechanisms.

We used $^{12}\rm{CO}(1-0)$ data from a Harvard--Smithsonian Center for Astrophysics (CfA) survey \citep{Dame2001ApJ...547..792D} to study the kinematics of the abovementioned molecular clouds. The velocity range for each molecular cloud is summarized in Table \ref{tab:cloud velocity range}, which was derived from molecular gas emission. The boundary of each molecular cloud was determined based on where the integrated intensity exceeds 3$\sigma$, with $\sigma$ representing the rms noise level.

\begin{table*}
	\centering
	\caption{The line of sight velocity range in the local standard of rest of the Gas in Each Molecular Cloud.}
	\begin{tabular}{cccccc}
		\hline
		Cloud & Orion~A & Orion~B & Perseus & Taurus & $\lambda$~Orionis \\ \hline
		$V_{\rm{LSR,rad}}$ & \multirow{2}*{[-2.6,18.2]} & \multirow{2}*{[-2.6,18.2]} & \multirow{2}*{[-5.2,15.6]} & \multirow{2}*{[-2.6,15.6]} & \multirow{2}*{[3.9,18.2]} \\
		(km s$^{-1}$) & & & & & \\ \hline
	\end{tabular}
	\label{tab:cloud velocity range}
\end{table*}

\subsection{Collecting and Selecting the Young Stellar Object Samples}

YSOs at different evolutionary stages were collected from 55 previous studies (see Table \ref{tab:ysotable}). For YSOs classified into uncertain or varied evolutionary stages by different catalogs, we assigned the classification that had the largest consensus. To acquire the parallax and proper motion for each object in the sample, we crossmatched them with the Gaia Data Release 3 (DR3) catalog. This comparison consisted of three situations: First, YSOs with Gaia EDR3/DR3 IDs were directly crossmatched using their \texttt{source\_id}. Second, YSOs with AllWISE IDs or Two Micron All Sky Surveuy IDs were matched with the \texttt{allwise\_best\_neighbour} catalog or the \texttt{tmass\_psc\_zsc\_best\_neighbour} catalog in the Gaia database, respectively. Third, for the remaining YSOs, we carried out a crossmatch between the YSOs and the Gaia DR3 catalog within a matching radius of 1$''$. Most of the YSOs we collected lack Gaia radial velocity measurements. Therefore, we added the radial velocities from APOGEE-2 DR17 \citep{Majewski2017}, crossmatched using their Gaia IDs. Finally, we obtained complete astrometric parameters and radial velocities for the YSOs ($\alpha, \delta, \varpi, \mu_\alpha^*, \mu_\delta$, and $v_r$).

\begin{table*}
	\centering
	\caption{Reference Young Stellar Object Catalogs For Each Molecular Cloud}
	\resizebox{\textwidth}{!}{
		\begin{tabular}{ccccc}
			\hline
			Orion~A & Perseus & Taurus & Orion~B & $\lambda$~Orionis\\ \hline
			\cite{Cottle2018} & \cite{Dunham2015} & \cite{Davies2014} & \cite{Flaherty2008} & \cite{Barrado2007} \\
			\cite{Davies2014} & \cite{Enoch2009} & \cite{Dent2013} & \cite{Furlan2016} & \cite{Barrado2011} \\
			\cite{Fang2013} & \cite{Evans2009} & \cite{Esplin2019} & \cite{Koenig2015} & \cite{Bayo2012} \\
			\cite{Furlan2016} & \cite{Gutermuth2008} & \cite{Gutermuth2009} & \cite{Kounkel2017} & \cite{Cao2022} \\
			\cite{Gross2019} & \cite{Gutermuth2009} & \cite{Hartmann2002} & \cite{Kryukova2012} & \cite{Koenig2015} \\
			\cite{Kim2013} & \cite{Kryukova2012} & \cite{Hartmann2005} & \cite{Marton2016} & \cite{Marton2016} \\
			\cite{Kim2016} & \cite{Lalchand2022} & \cite{Joncour2017} & \cite{Megeath2012} & \\
			\cite{Kounkel2019} & \cite{Marton2016} & \cite{Kraus2017} & \cite{Principe2014} & \\
			\cite{Kryukova2012} & \cite{Mercimek2017} & \cite{Lopez2021} & \cite{Spezzi2015} & \\
			\cite{Marton2016} & \cite{Meuench2007} & \cite{Luhman2010} &  & \\
			\cite{Megeath2012} & \cite{Ruiz2018} & \cite{Marton2016} &  & \\
			\cite{Minier2003} & \cite{Tobin2016} & \cite{Rebull2010} &  & \\
			\cite{Pillitteri2013} & \cite{Young2015} & \cite{Rebull2020} &  & \\
			&  & \cite{White2004} &  & \\ \hline
	\end{tabular}}
	\label{tab:ysotable}
\end{table*}

YSOs with a renormalized unit weight error (\texttt{ruwe}) $\geq 1.4$ \citep{Lindegren2021} or \texttt{parallax/parallax\_error} $\leq 5$ were excluded due to poor astrometric solutions or imprecise parallax measurements. Furthermore, to ensure precise kinematic information, strict criteria were applied to the uncertainties of the proper motion and radial velocity: $err_{\mu_{\alpha^*}}, err_{\mu_\delta} < 1$ mas yr$^{-1}$, and \texttt{VERR} $<$ 1 km s$^{-1}$ \citep{Gro2021A&A...647A..91G}. Gaia, being an optical telescope, lacks sensitivity to high-extinction sources, thus missing Class 0 and I YSOs that are deeply embedded in clouds or surrounded by dust envelopes. Consequently, our sample contains no Class 0 YSOs and only a few Class I YSOs, and the study mainly considers Class II and III YSOs. These YSO samples are accessible in the supplementary materials. We have collected the \texttt{mass\_flame} from Gaia DR3 catalog and the spectral types from SIMBAD\footnote{\url{http://simbad.u-strasbg.fr/simbad/}} available for YSO samples. Additionally, we further split each molecular cloud into different subregions based on previous studies, see details in Section \ref{results}. Hence, different parallax criteria were adopted to select YSOs in each subregion. For each subregion, we calculated the mean parallax of the YSOs and used $\pm 10\%$ around this value as the parallax limit, as listed in Table \ref{tab:parallax limit} (see Appendix \ref{sec:appendix} for the distribution of the YSO samples in each subregion). 

\begin{table*}
	\centering
	\caption{Parallax Criteria for the Young Stellar Object Samples within Each Subregion}
	\begin{tabular}{ccccccc}
		\hline
		\multirow{3}*{Orion~A}& Subregion & OMC & L1641N & L1641C & L1641S1 & L1641S2  \\ 
		\cline{2-7}&parallax limit & \multirow{2}*{[2.31,2.81]} & \multirow{2}*{[2.37,2.80]} & \multirow{2}*{[2.33,2.77]} & \multirow{2}*{[2.21,2.61]} & \multirow{2}*{[2.22,2.61]}  \\
		&(mas) & & & & & \\ \hline
		\multirow{3}*{Orion~B}& Subregion & L1630N & NGC 2024 &  &  &  \\ 
		\cline{2-7}&parallax limit & \multirow{2}*{[2.10,2.55]} & \multirow{2}*{[2.35,2.8]} &  &  &   \\
		&(mas) & & & & & \\ \hline
		\multirow{3}*{Perseus}& Subregion & IC 348 & NGC 1333 &  &  &   \\ 
		\cline{2-7}&parallax limit & \multirow{2}*{[3.04,3.71]} & \multirow{2}*{[3.11,3.59]} &  &  &   \\
		&(mas) & & & & & \\ \hline	
		\multirow{3}*{Taurus}& Subregion & Heiles' Cloud 2 & L1495 & L1536 & B18 & B213  \\ 
		\cline{2-7}&parallax limit & \multirow{2}*{[6.42,7.84]} & \multirow{2}*{[7.00,8.18]} & \multirow{2}*{[5.55,6.79]} & \multirow{2}*{[7.00,8.56]} & \multirow{2}*{[5.66,6.91]}  \\
		&(mas) & & & & &  \\ \hline
		\multirow{3}*{$\lambda$~Orionis}& Subregion & B30 & &  &  &   \\ 
		\cline{2-7}&parallax limit & \multirow{2}*{[2.36,2.64]} &  &  &  &   \\
		&(mas) & & & & & \\ \hline
	\end{tabular}
	\label{tab:parallax limit}
\end{table*}

\subsection{Velocity Determination for the Young Stellar Objects and Clouds}

To compare the radial velocities of the YSOs and molecular clouds, we converted the heliocentric radial velocity ($V_{\rm{HELIO\_AVG}}$) of each YSO into a local standard of rest (LSR) velocity ($V_{\rm{LSR,rad}}$) \citep{Kerr1986}. The 3D velocities ($V_{\rm{3D}}=\sqrt{\rm{V_x}^2+\rm{V_y}^2+\rm{V_z}^2}$) of the YSOs in the Cartesian coordinate system were calculated using their astrometric positions ($\alpha, \delta$), parallaxes ($\varpi$), proper motions ($\mu_{\alpha*}, \mu_\delta$), and $V_{\rm{LSR,rad}}$. $\rm{V_x}$, $\rm{V_y}$, and $\rm{V_z}$ are the velocity components the same direction as Galactic rotation but perpendicular to the Sun–-Galaxy center line, in the direction of the Galactic center to toward the Sun, and towards the north Galactic pole, respectively. In each subregion, we took the mean values of the LSR and 3D velocities for the Class II and III YSOs, respectively, and the standard deviation as the velocity dispersion $\sigma_{\rm{YSO}}$. The errors were calculated using $e_{\rm{YSO}}=\sigma_{\rm{YSO}}/\sqrt{\rm{N}}$, where $\rm{N}$ is the number of YSOs. 

The difference in the LSR velocity between the YSOs and molecular gas can be used to investigate how stars emerge from their natal molecular cloud, i.e., $|\Delta V_{\rm{LSR,rad}}|=|V_{\rm{LSR,rad,gas}}-\overline{V_{\rm{LSR,rad,YSO}}}|$, where $\overline{V_{\rm{LSR,rad,YSO}}}$ represents the mean LSR velocity of the YSOs in each subregion. The kinematic evolution of the YSOs under stellar feedback can be constrained by comparing the 1D and 3D kinematic characteristics between the Class II and III YSOs, i.e., for 1D, $|\Delta V_{\rm{1D}}|=|\overline{V_{\rm{LSR,rad,III}}}-\overline{V_{\rm{LSR,rad,II}}}|$, where $\overline{V_{\rm{LSR,rad,II}}}$ and $\overline{V_{\rm{LSR,rad,III}}}$ represent the mean LSR velocities of the Class II and III YSOs in each subregion, respectively. And for 3D, $|\Delta V_{\rm{3D}}|=|\overline{V_{\rm{3D,III}}}-\overline{V_{\rm{3D,II}}}|$, where $\overline{V_{\rm{3D,II}}}$ and $\overline{V_{\rm{3D,III}}}$ represent the mean 3D velocities of the Class II and III YSOs in each subregion, respectively.

\section{Results}
\label{results}

In this section we describe the environments and major stellar feedback activities such as outflows, H~\textsc{ii} regions, and supernovae in every cloud. We also present the distributions of the YSO samples, as well as their proper motions and radial velocities. Finally, we present the $|\Delta V_{\rm{LSR,rad}}|$ values of the Class II and III YSOs, and the $|\Delta V_{\rm{1D}}|$ and $|\Delta V_{\rm{3D}}|$ values in every subregion.

\subsection{Orion~A}

\begin{figure}
	\centering
	\includegraphics[width=0.8\linewidth]{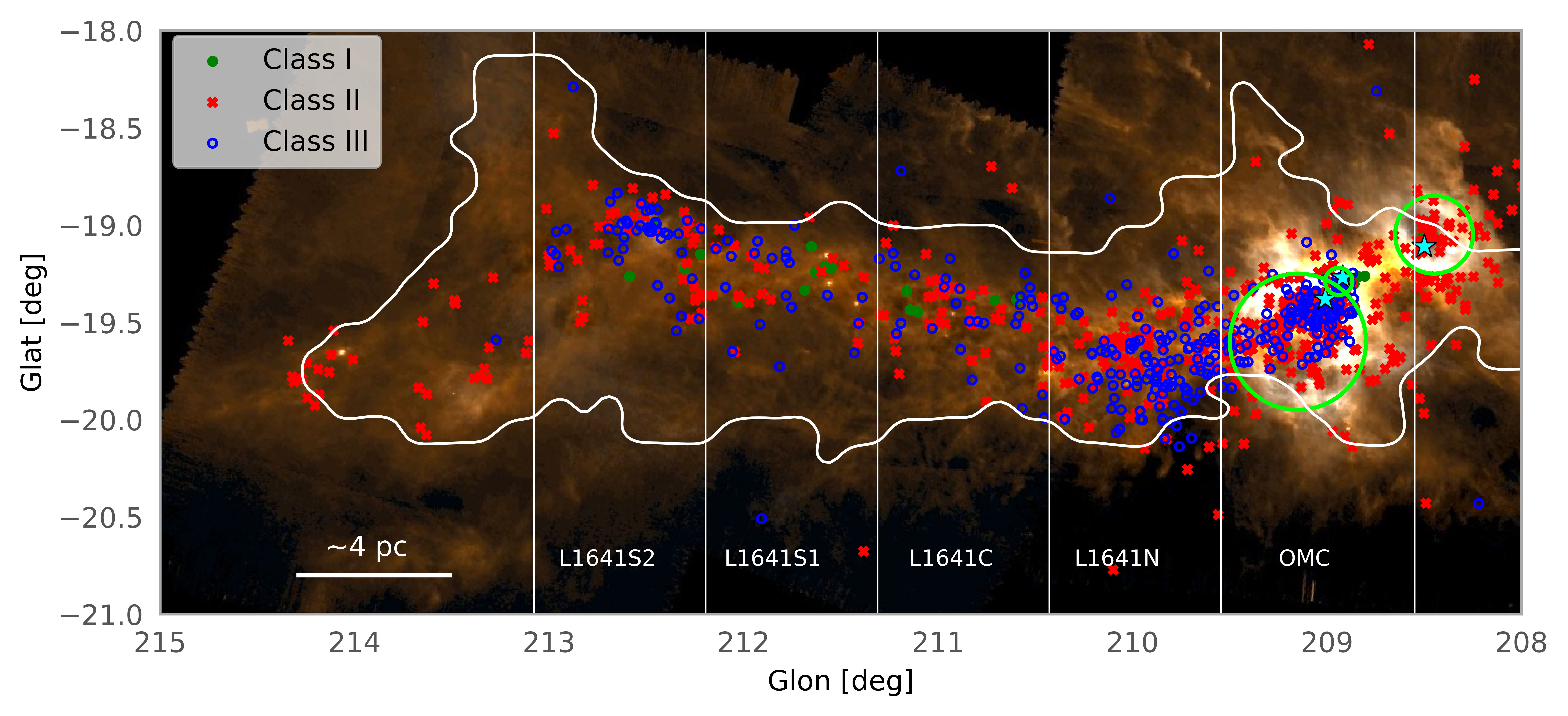}
	\caption{Projected distributions of the Class I (green filled circles), Class II (red crosses), and Class III (blue hollow circles) YSOs in Orion~A. The background is a false-color image of Orion~A observed by Herschel's Photodetector Array Camera and Spectrometer at 70 $\mu$m (blue) and 160 $\mu$m (red). The three cyan stars from left to right represent $\theta^1$ Ori~C, NU~Ori, and 42~Orionis, respectively. The green circles from left to right represent the H~\textsc{ii} regions M42, M43, and NGC~1977, respectively. The contour represents the boundary of the cloud, which has a velocity range of [$-$2.6, 18.2] km s$^{-1}$.}
	\label{fig:orionapicyso}
\end{figure}

Orion~A is a cometary molecular cloud complex at a distance of $\sim$ 414 pc \citep{Menten2007}, consisting of a denser and more active star forming head, the Orion Molecular Cloud (OMC), and a less dense and relatively quiescent tail formed of L1641 and L1647 \citep[e.g.][]{Allen2008, Muench2008}. L1641 spans most of the Orion~A molecular cloud, roughly 6.3 deg$^2$. In this study, we focus on the OMC and L1641 in Orion~A, where L1641 is split into four subregions, i.e., L1641N, L1641C, L1641S1, and L1641S2, following the work by \citet{Gro2021A&A...647A..91G}. These subregions are shown in Figure \ref{fig:orionapicyso}.

Feedback in the OMC is dominated by photoionization, radiation pressure, and hot stellar winds from massive stars. OMC-1 is the most massive core in the OMC, containing the massive O7V-type star $\theta^1$ Ori~C and the associated H~\textsc{ii} region , M42 \citep[e.g.][]{Pabst2019Natur.565..618P, Pabst2020}, and two smaller expanding bubbles, i.e., M43 and NGC~1977, driven by a B0.5V star (NU~Ori) and a B1V star (42~Orionis), respectively \citep{Pabst2020}. \citet{Pabst2019Natur.565..618P} indicated that the stellar winds of $\theta^1$ Ori~C have already swept away the surrounding material, forming a "bubble" with a diameter of about 4 pc and whose shell is expanding at a speed of 13 km s$^{-1}$.

L1641 contains abundant outflows and Herbig--Haro objects \citep{Allen2008}, with outflows being the primary form of stellar feedback. L1641N is located in the northern part of L1641 and contains several young star clusters \citep{Gomez1998}. Optical, X-ray, and infrared studies suggest that most of the protostars clustered in this region have assembled into small clusters or groups containing a few dozen YSOs \citep{Strom1993}.

\begin{table*}
	\centering
	\caption{Kinematic Information of the Young Stellar Objects in Orion~A}
	\label{tab:OrionA info}
	\begin{tabular}{cccccccccccc}
		\toprule
		\multirow{2}{*}{region} & \multirow{2}{*}{Class} & \multirow{2}{*}{N} & $\mu_{\alpha^*}$ & $\mu_\delta$ & $V_{\rm{LSR,rad}}$ & $\sigma_{V_{\rm{LSR,rad}}}$ & $|\Delta V_{\rm{LSR,rad}}|$ & $|\Delta V_{\rm{1D}}|$ & $V_{\rm{3D}}$ & $\sigma_{V_{\rm{3D}}}$ & $|\Delta V_{\rm{3D}}|$ \\
		& & & mas yr$^{-1}$ & mas yr$^{-1}$ & km s$^{-1}$ & km s$^{-1}$ & km s$^{-1}$ & km s$^{-1}$ & km s$^{-1}$ & km s$^{-1}$ & km s$^{-1}$ \\ \hline
		
		\multirow{4}{*}{OMC} & Cloud & - & - & - & 8.9 & 1.1 & - & - & - & -& - \\ 
		\cline{2-12} & I & 12 & 0.98$\pm$0.07 & -0.74$\pm$0.10 & 11.6$\pm$0.7 & 2.5 & 2.7$\pm$0.7 & - & 9.7$\pm$0.7 & 2.3 & - \\ 
		\cline{2-12} & II & 228 & 1.35$\pm$0.01 & -0.07$\pm$0.01 & 8.5$\pm$0.2 & 2.4 & 0.4$\pm$0.2 & - & 7.3$\pm$0.1 & 2.0 & - \\ 
		\cline{2-12} & III & 105 & 1.29$\pm$0.01 & 0.20$\pm$0.01 & 9.2$\pm$0.3 & 3.1 & 0.3$\pm$0.3 & 0.7$\pm$0.4 & 8.0$\pm$0.2 & 2.5 & 0.7$\pm$0.2 \\ \hline
		
		\multirow{4}{*}{L1641N}& Cloud & - & - & - & 7.7 & 0.7 & - & - & - & -& - \\ 
		\cline{2-12} & I & 1 & -0.37$\pm$0.44 & 0.31$\pm$0.38 & 5.8$\pm$0.1 & - & 1.9$\pm$0.1 & - & 4.7$\pm$0.3 & - & - \\ 
		\cline{2-12} & II & 76 & 1.00$\pm$0.01 & 0.15$\pm$0.01 & 7.5$\pm$0.3 & 2.4 & 0.2$\pm$0.3 & - & 6.2$\pm$0.2 & 2.1 & - \\ 
		\cline{2-12} & III & 99 & 1.05$\pm$0.01 & 0.23$\pm$0.01 & 8.3$\pm$0.3 & 2.9 & 0.6$\pm$0.3 & 0.8$\pm$0.4 & 7.2$\pm$0.2 & 2.4 & 1.0$\pm$0.3 \\ \hline
		
		\multirow{4}{*}{L1641C}& Cloud & - & - & - & 6.5 & 0.9 & - & - & - & - & - \\ 
		\cline{2-12} & I & 5 & 0.45$\pm$0.10 & -0.41$\pm$0.13 & 8.6$\pm$2.2 & 4.9 & 2.1$\pm$2.2 & - & 6.5$\pm$2.1 & 4.8 & - \\ 
		\cline{2-12} & II & 38 & 0.37$\pm$0.02 & -0.47$\pm$0.02 & 5.3$\pm$0.4 & 2.7 & 1.2$\pm$0.4 & - & 3.6$\pm$0.4 & 2.7 & - \\ 
		\cline{2-12} & III & 17 & 0.77$\pm$0.07 & -0.27$\pm$0.07 & 5.4$\pm$0.7 & 2.8 & 1.1$\pm$0.7 & 0.1$\pm$0.8 & 4.5$\pm$0.7 & 3.0 & 0.9$\pm$0.8 \\ \hline
		
		\multirow{4}{*}{L1641S1}& Cloud & - & - & - & 6.1 & 0.8 & - & - & -&-& - \\ 
		\cline{2-12} & I & 8 & -0.17$\pm$0.09 & -0.63$\pm$0.09 & 6.8$\pm$1.4 & 3.8 & 0.7$\pm$1.4 &-& 4.6$\pm$1.0 & 3.0 & - \\  
		\cline{2-12} & II & 14 & 0.22$\pm$0.04 & -0.60$\pm$0.04 & 4.9$\pm$0.5 & 1.9 & 1.2$\pm$0.5 &-& 2.9$\pm$0.5 & 1.7 & -  \\ 
		\cline{2-12} & III & 12 & 0.24$\pm$0.04 & -0.20$\pm$0.06 & 6.0$\pm$0.6 & 2.0 & 0.1$\pm$0.6 &1.1$\pm$0.8& 3.8$\pm$0.6 & 2.1 & 0.9$\pm$0.8 \\ \hline
		
		\multirow{4}{*}{L1641S2}& Cloud & - & - & - & 5.2 & 0.9 & - & - & -&-& - \\ 
		\cline{2-12} & I & 6 & 0.32$\pm$0.11 & -0.49$\pm$0.12 & 4.1$\pm$0.5 & 1.2 & 1.1$\pm$0.5 &-& 2.8$\pm$0.5 & 1.2 & - \\ 
		\cline{2-12} & II & 39 & 0.38$\pm$0.01 & -0.35$\pm$0.01 & 4.3$\pm$0.4 & 2.8 & 0.9$\pm$0.4 &-& 2.8$\pm$0.4 & 2.3 & - \\ 
		\cline{2-12} & III & 28 & 0.32$\pm$0.02 & -0.47$\pm$0.02 & 4.6$\pm$0.7 & 3.8 & 0.6$\pm$0.7 &0.3$\pm$0.8& 3.4$\pm$0.6 & 3.1 & 0.6$\pm$0.7 \\ 
		\bottomrule
	\end{tabular}
\end{table*}

Previous studies have revealed that the Orion~A molecular cloud has a distance gradient from head to tail \citep{Gross2018}; thus, we applied different parallax criteria to the YSO samples of the five subregions (see the parallax criteria listed in Table \ref{tab:parallax limit}). Following this strategy, we obtained 688 YSOs with precise parallax and 3D kinematic information in the Orion~A region, comprising 32 Class I, 395 Class II, and 261 Class III YSOs. The number of YSOs in each subregion is listed in Table \ref{tab:OrionA info}.

Figure \ref{fig:orionapicyso} shows the distribution of all YSOs in Orion~A, aligned along the molecular cloud. Most of the YSOs (approximately 76\%) are located in the OMC and L1641N, consistent with the findings of \citet{Gross2019} and \citet{Megeath2012}. In the OMC, most of the YSOs are concentrated around the H~\textsc{ii} regions M42 and M43, particularly near the massive star $\theta^{1}$ Ori~C. These results imply that the environments in the tail region of Orion~A are distinct from those near the head region, with the OMC being the most active subregion of star formation.

\begin{figure}
	\centering
	\includegraphics[scale=0.6]{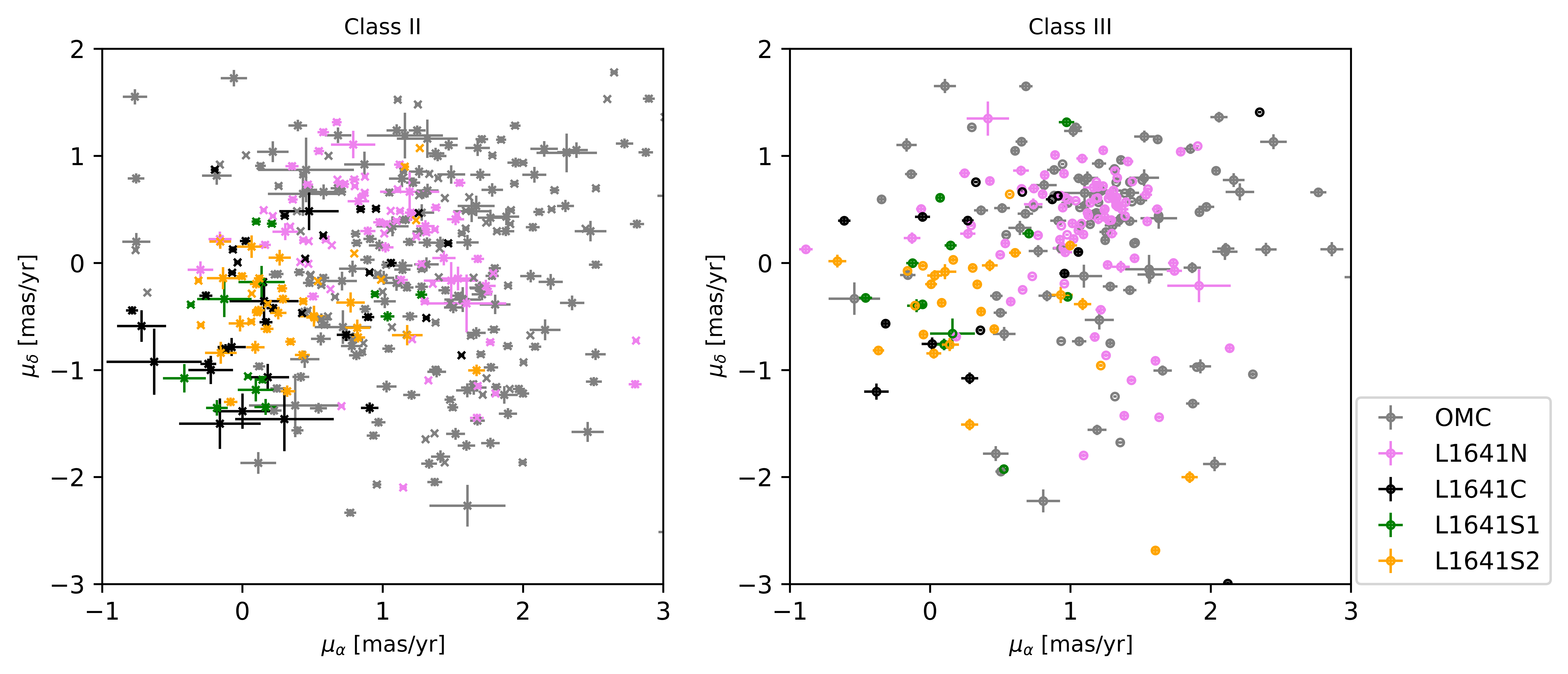}
	\caption{Distribution of the proper motions of the YSOs in the subregions of Orion~A, where the crosses and circles represent Class II and III YSOs, respectively.}
	\label{fig:orionapm}
\end{figure}

\begin{figure}
	\centering
	\includegraphics[scale=0.7]{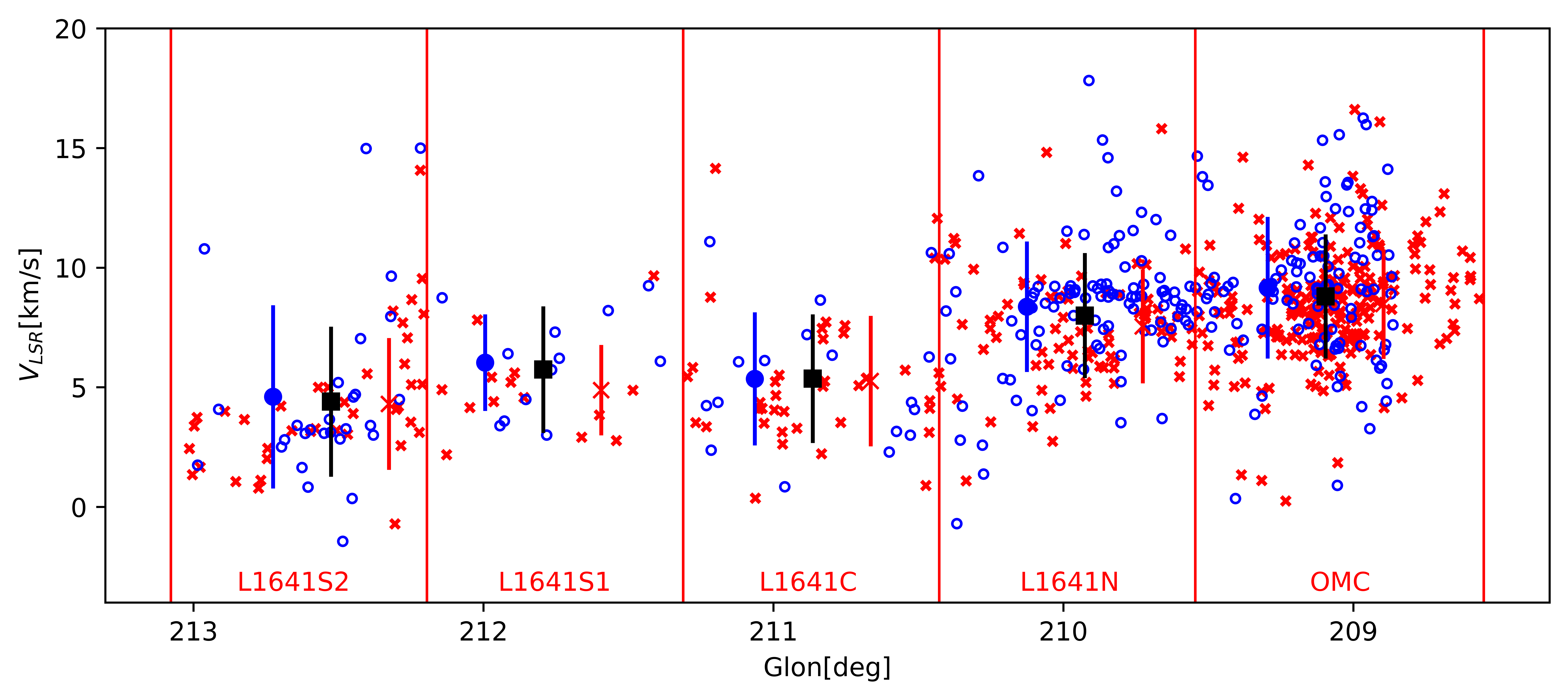}
	\caption{Position--velocity (PV) diagram for the Class II (red crosses) and Class III (blue circles) YSOs in Orion~A. The average velocity per region for all YSOs, as well as for the Class II and III YSOs individually are shown with black squares, red crosses, and blue circles, respectively, with the error bars representing the velocity dispersion per region.}
	\label{fig:orialvplot}
\end{figure}

The distribution of YSO proper motions in the five subregions is presented in Figure \ref{fig:orionapm}. In the OMC and L1641N, the Class III YSOs exhibit more concentrated proper motions compared to the Class II YSOs. However, in the other three subregions (i.e., L1641C, L1641S1, and L1641S2), the Class II YSOs show more concentrated proper motions than the Class III YSOs. A PV diagram of the YSOs in Orion~A is shown in Figure \ref{fig:orialvplot}. The YSOs at the head of Orion~A display larger LSR velocities compared to those in the tail region. The average LSR velocities of the YSOs in the OMC and L1641N are similar, about 8.4 km s$^{-1}$. In contrast, the other three subregions have similar LSR velocities, with an average of about 5.1 km s$^{-1}$. The average proper motions and LSR velocities of the YSOs in each subregion are listed in Table \ref{tab:OrionA info}. The results of a previous analysis and the kinematic information listed in Table \ref{tab:OrionA info} indicate that the OMC and L1641N exhibit similar kinematic characteristics, and the three subregions in the tail (i.e., L1641C, L1641S1, and L1641S2) show similar kinematic characteristics. 

Table \ref{tab:OrionA info} also summarizes the kinematics of the YSOs in Orion~A, including the LSR velocity offset between the YSOs and the parent molecular cloud ($|\Delta V_{\rm{LSR,rad}}|$) and the difference in the 3D velocities between the Class II and III YSOs ($|\Delta V_{\rm{3D}}|$). Except for L1641N, the Class II YSOs have larger $|\Delta V_{\rm{LSR,rad}}|$ than the Class III YSOs. The YSOs in L1641C exhibit the largest LSR velocity offsets from the parent molecular cloud, with 1.2 km s$^{-1}$ for the Class II and 1.1 km s$^{-1}$ for the Class III YSOs. Additionally, the largest difference in $|\Delta V_{\rm{LSR,rad}}|$ between the Class II and III YSOs occurs in L1641S1, 1.1 km s$^{-1}$, while it does not exceed 0.5 km s$^{-1}$ in the other subregions. The minimum and maximum $|\Delta V_{\rm{1D}}|$ values are 0.1 km s$^{-1}$ in L1641C and 1.1  km s$^{-1}$ in L1641S1, respectively. In the OMC and L1641S2, $|\Delta V_{\rm{3D}}|$ is 0.7 and 0.6 km s$^{-1}$, respectively, while in the other subregions, these values are close to 1 km s$^{-1}$. The YSOs in the OMC and L1641N exhibit similar 3D velocities and dispersions, and those in the other three subregions are similar. The 3D velocity dispersion of the Class II YSOs is smaller than that of the Class III YSOs in all five subregions. 

\subsection{Orion~B}

Orion~B is located approximately 400 pc from the Sun, with a total mass of about $7\times10^4$ M$_\odot$ \citep{Lombardi2014A&A...566A..45L}. The star formation activity in Orion~B is primarily concentrated in NGC~2024 and L1630N, which are located in the northern and southern parts of Orion~B, respectively, and contain 60\%--90\% of the YSOs in this region \citep{Meyer2008hsf1.book..662M}. NGC~2024 is a representative massive star-forming region, with its morphology in the optical band being dominated by a north--south oriented dark belt of high optical depth, which is likely a foreground dust layer. NGC~2024 hosts two massive stars: a highly reddened late O8-type star, IRS~2b \citep{Bik2003} and an O9.71b-type star, $\zeta$~Ori \citep{Hummel2013}. IRS~2b is situated on the eastern side of the dark belt and may be the ionization source of the H~\textsc{ii} region, while $\zeta$~Ori is located on the south side of NGC~2024. L1630N comprises the bright optical reflection nebulae NGC~2068 and NGC~2071. Star formation activity is still ongoing in this region \citep{Spezzi2015}, with outflows being the main type of stellar feedback. Other parts of the Orion~B are characterized by vast and relatively quiet regions with numerous filamentary structures \citep{Orkisz2017A&A...599A..99O}.

\begin{table*}
	\centering
	\caption{Kinematic Information of the Young Stellar Objects in Orion~B}
	\label{tab:OrionB info}
	\begin{tabular}{cccccccccccc}
		\hline
		\multirow{2}{*}{region} & \multirow{2}{*}{Class} & \multirow{2}{*}{N} & $\mu_{\alpha^*}$ & $\mu_\delta$ & $V_{\rm{LSR,rad}}$ & $\sigma_{V_{\rm{LSR,rad}}}$ & $|\Delta V_{\rm{LSR,rad}}|$&$|\Delta V_{\rm{1D}}|$ & $V_{\rm{3D}}$ & $\sigma_{V_{\rm{3D}}}$ & $|\Delta V_{\rm{3D}}|$ \\
		 & & & mas yr$^{-1}$ & mas yr$^{-1}$ & km s$^{-1}$ & km s$^{-1}$ & km s$^{-1}$& km s$^{-1}$ & km s$^{-1}$ & km s$^{-1}$ & km s$^{-1}$ \\ \hline
		\multirow{4}{*}{L1630N} & Cloud & - & - & - & 9.3 & 0.8 & - & - &- & - & - \\ 
		\cline{2-12} & I & 0 & - & - & - & - & - & - &- & - & - \\ 
		\cline{2-12} & II & 42 & -0.68$\pm$0.02 & -0.88$\pm$0.01 & 10.7$\pm$0.1 & 2.0 & 1.4$\pm$0.3 &- & 8.6$\pm$0.1 & 1.9 & - \\ 
		\cline{2-12} & III & 13 & 0.09$\pm$0.06 & -0.89$\pm$0.02 & 11.2$\pm$0.1 & 1.7 & 1.9$\pm$0.3 &0.5$\pm$0.1 & 9.1$\pm$0.1 & 1.6 & 0.5$\pm$0.1 \\ \hline
		\multirow{4}{*}{NGC 2024}& Cloud & - & - & - & 10.0 & 0.4 & - & - &- & - & - \\ 
		\cline{2-12} & I & 0 & - & - & - & - & - & - & - &- & - \\ 
		\cline{2-12} & II & 36 & 0.24$\pm$0.02 & -0.59$\pm$0.03 & 9.9$\pm$0.1 & 2.9 & 0.1$\pm$0.2 &- & 8.7$\pm$0.1 & 2.0 & - \\ 
		\cline{2-12} & III & 4 & 1.27$\pm$0.18 & -1.23$\pm$0.01 & 8.7$\pm$1.1 & 4.3 & 1.3$\pm$1.1 &1.2$\pm$1.1 & 7.3$\pm$0.9 & 3.7 & 1.4$\pm$0.9 \\ \hline
	\end{tabular}
\end{table*}

The distance to L1630N and NGC~2024 is 400 pc \citep{Gibb2008} and 420 pc \citep{Kounkel2017}, respectively. We employed different parallax criteria to select YSOs within these two subregions, i.e., [2.10, 2.55] mas and [2.35, 2.87] mas, respectively (see Table \ref{tab:parallax limit} and subregions in Figure \ref{fig:orionbpicyso}). The parallax distributions of the YSOs in these regions are presented in Appendix \ref{sec:appendix}. 95 YSOs in the Orion~B region with precise distances and 3D kinematics measurements were retrieved, comprising 78 Class II and 17 Class III YSOs. The number of YSOs in each subregion is listed in Table \ref{tab:OrionB info}.

\begin{figure}
	\centering
	\includegraphics[width=0.5\linewidth]{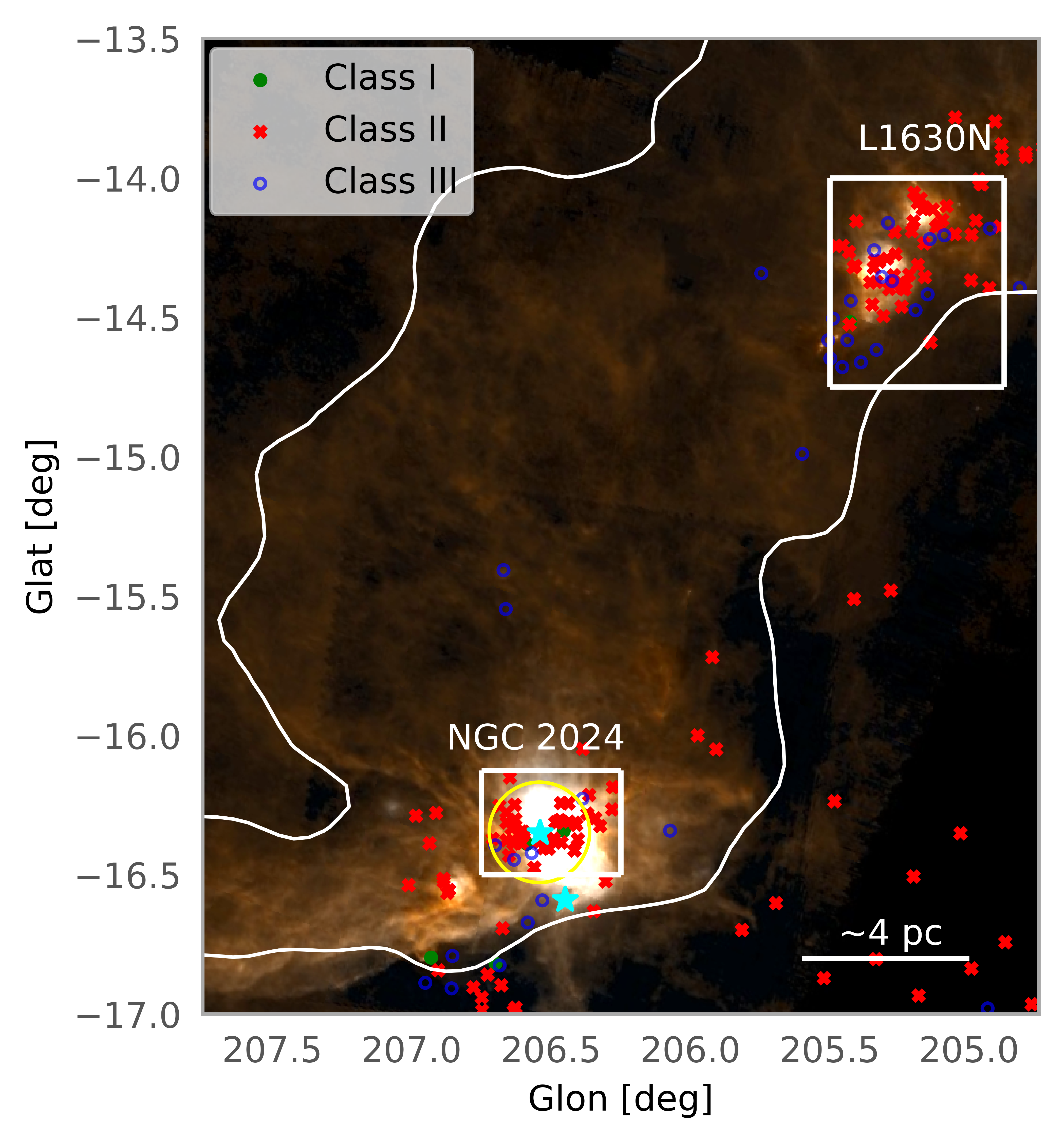}
	\caption{Projected distribution of YSOs in Orion~B. The yellow circle represents the H~\textsc{ii} region driven by IRS~2b. The cyan stars represent IRS~2b and the massive O-type star $\zeta$~Ori, from north to south. L1630N, from top to bottom, comprises NGC~2071 and NGC~2068. The contours represent the velocity channel of [$-$2.6, 18.2] km s$^{-1}$. The other features are the same in Figure \ref{fig:orionapicyso}.}
	\label{fig:orionbpicyso}
\end{figure}

The projected distribution of YSOs in Orion~B are shown in Figure \ref{fig:orionbpicyso}. The Class II YSOs in L1630N are mainly located around two reflection nebulae, NGC~2071 and NGC~2068, while in NGC~2024 the YSOs are almost exclusively located around the H~\textsc{ii} region. Both L1630N and NGC~2024 have fewer Class III YSOs compared to Class II, with the Class III YSOs accounting for approximately 18\% of the total population. This may be due to the relative youth of these two subregions, as the age of NGC~2024 is estimated to be 0.5 Myr \citep{Haisch2000}, and the age of L1630N might be 1--2 Myr \citep[e.g.][]{Flaherty2008,Spezzi2015}. 

\begin{figure}
	\centering
	\includegraphics[scale=0.6]{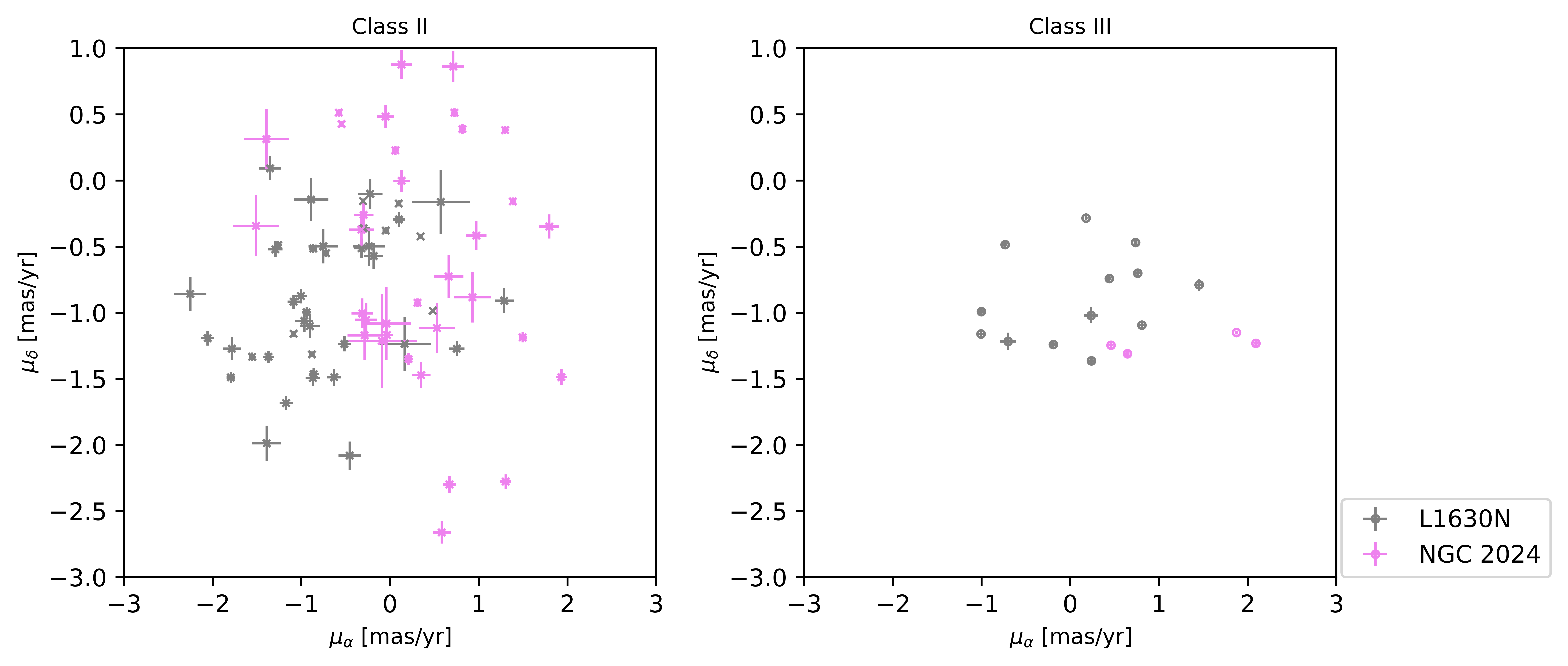}
	\caption{Proper motion distribution of YSOs in Orion~B, where the crosses represent Class II YSOs and the circles represent Class III YSOs.}
	\label{fig:orionbpm}
\end{figure}

The proper motion distribution of the YSOs in L1630N and NGC~2024 is presented in Figure \ref{fig:orionbpm}. The Class II and III YSOs in L1630N show similar proper motions in the decl. direction, with a difference of 0.77 mas yr$^{-1}$ (1.5 km s$^{-1}$) in the R.A. direction.

Table \ref{tab:OrionB info} presents the kinematics of the YSOs in L1630N and NGC~2024, including the LSR velocity offsets of the YSOs from the parent molecular cloud and the 3D velocities of the YSOs. In L1630N, both the Class II and III YSOs show relatively large offsets from the parent cloud in the line-of-sight direction, i.e., 1.4 km s$^{-1}$ and 1.9 km s$^{-1}$, respectively. The $|\Delta V_{\rm{1D}}|$ value in L1630N is 0.5 km s$^{-1}$. The 3D velocities of the Class II YSOs are smaller than those of the Class III YSOs, with the difference between them being about 0.5 km s$^{-1}$. The 3D velocity dispersion of the Class II YSOS is larger than that of the Class III YSOs. In NGC~2024, the LSR velocity difference between the parent cloud and the Class II YSOs is only 0.1 km s$^{-1}$. Due to the lack of Class III YSOs in this region, no further analysis of the kinematic characteristics of the Class III YSOs was performed.

\subsection{Perseus}

The Perseus molecular cloud complex is at a distance of about 270 pc \citep{Zucker2020A&A...633A..51Z}, with a total mass of about $7\times10^3$ M$_\odot$ and a total area of approximately 70 pc$^2$. According to \citet{Sancisi1974}, Perseus has been influenced by the final stages of a supernova that occurred several million years ago in the Per OB2 association. The YSOs in Perseus are primarily concentrated in two regions: the intermediate-mass star-forming region NGC~1333, known for its numerous outflows \citep[e.g.][]{ Arce2010ApJ...715.1170A,Plunkett2013, Dionatos2017}, and the young open cluster IC~348, with member stars between 0.5 and 10 Myr old, most of which are younger than 3 Myr \citep[e.g.][]{Luhman1998,Meuench2007, Stelzer2012}.

\begin{table*}
	\centering
	\caption{Kinematic Information of the Young Stellar Objects in Perseus}
	\label{tab:Perseus info}
	\begin{tabular}{cccccccccccc}
		\hline
		\multirow{2}{*}{region} & \multirow{2}{*}{Class} & \multirow{2}{*}{N} & $\mu_{\alpha^*}$ & $\mu_\delta$ & $V_{\rm{LSR,rad}}$ & $\sigma_{V_{\rm{LSR,rad}}}$ & $|\Delta V_{\rm{LSR,rad}}|$&$|\Delta V_{\rm{1D}}|$ & $V_{\rm{3D}}$ & $\sigma_{V_{\rm{3D}}}$ & $|\Delta V_{\rm{3D}}|$ \\
		& & & mas yr$^{-1}$ & mas yr$^{-1}$ & km s$^{-1}$ & km s$^{-1}$ & km s$^{-1}$& km s$^{-1}$ & km s$^{-1}$ & km s$^{-1}$ & km s$^{-1}$ \\ \hline
		\multirow{4}{*}{IC 348} & Cloud & - & - & - & 8.4 & 0.7 & - & - &- & - & - \\ 
		\cline{2-12} & I & 0 & - & - & - & - & - & - &- & - & - \\ 
		\cline{2-12} & II & 62 & 4.61$\pm$0.01 & -6.63$\pm$0.01 & 8.5$\pm$0.2 & 1.6 & 0.1$\pm$0.2 &- & 10.1$\pm$0.2 & 1.4 & - \\ 
		\cline{2-12} & III & 29 & 4.63$\pm$0.02 & -6.22$\pm$0.02 & 9.5$\pm$0.5 & 2.8 & 1.1$\pm$0.5 &1.0$\pm$0.5 & 11.2$\pm$0.5 & 2.6 & 1.1$\pm$0.5 \\ \hline
		\multirow{4}{*}{NGC 1333} & Cloud & - & - & - & 6.9 & 0.3 & - & - &- & - & - \\ 
		\cline{2-12} & I & 2 & 7.98$\pm$0.63 & -10.36$\pm$0.31 & 8.5$\pm$0.3 & 0.5 & 1.6$\pm$0.3 &- & 11.1$\pm$1.3 & 1.9 & - \\ 
		\cline{2-12} & II & 27 & 7.30$\pm$0.03 & -9.97$\pm$0.02 & 7.8$\pm$0.4 & 1.9 & 0.9$\pm$0.5 &- & 9.6$\pm$0.3 & 1.7 & - \\ 
		\cline{2-12} & III & 15 & 5.75$\pm$0.23 & -10.28$\pm$0.05 & 7.4$\pm$1.0 & 3.7 & 0.5$\pm$1.0 &0.4$\pm$1.1 & 10.3$\pm$0.6 & 2.4 & 0.7$\pm$0.7 \\ \hline
	\end{tabular}
\end{table*}

\begin{figure}
	\centering
	\includegraphics[scale=0.7]{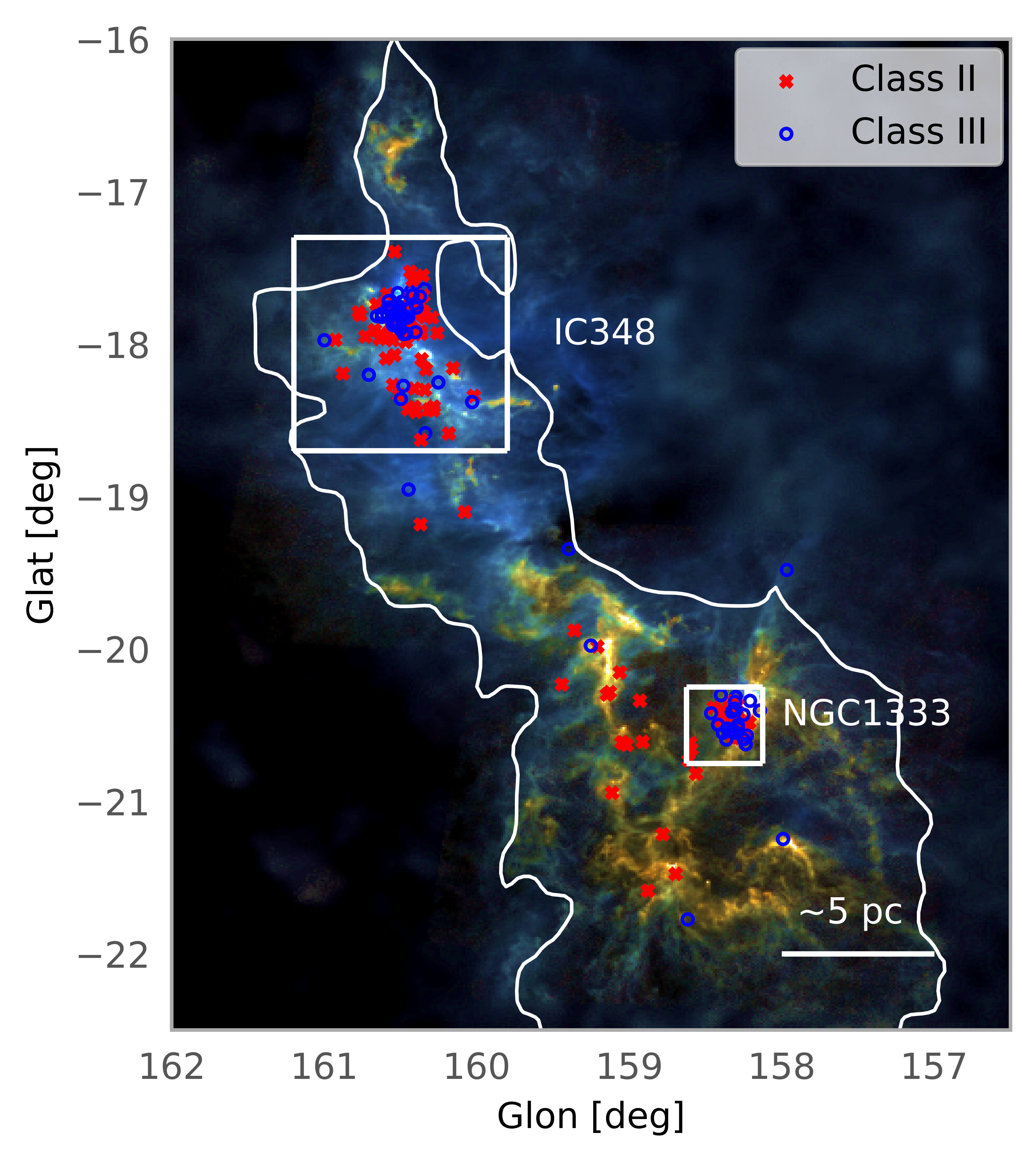}
	\caption{Projected distribution of YSOs in Perseus. The background is a false-color image of Perseus observed by Herschel/SPIRE at 250 $\mu$m (blue), 350 $\mu$m (green) and 500 $\mu$m (red). The contours represent the velocity channel of [$-$5.2, 15.6] km s$^{-1}$. The other features are the same as in Figure \ref{fig:orionapicyso}.}
	\label{fig:perseuspicyso}
\end{figure}

IC~348 and NGC~1333 are located at a distance of 321 pc and 293 pc \citep{Ortiz2018}, respectively. These two subregions are shown in Figure \ref{fig:perseuspicyso}. Hence we applied different parallax criteria to select YSOs in each subregion (see Table \ref{tab:parallax limit}), where the parallax distribution of each subregion is shown in Appendix \ref{sec:appendix}. Finally, 135 YSOs in Perseus with precise distances and 3D kinematics measurements were selected, comprising two Class I, 89 Class II, and 44 Class III YSOs. The number of YSOs in each subregion is listed in Table \ref{tab:Perseus info}.

\begin{figure}
	\centering
	\includegraphics[scale=0.6]{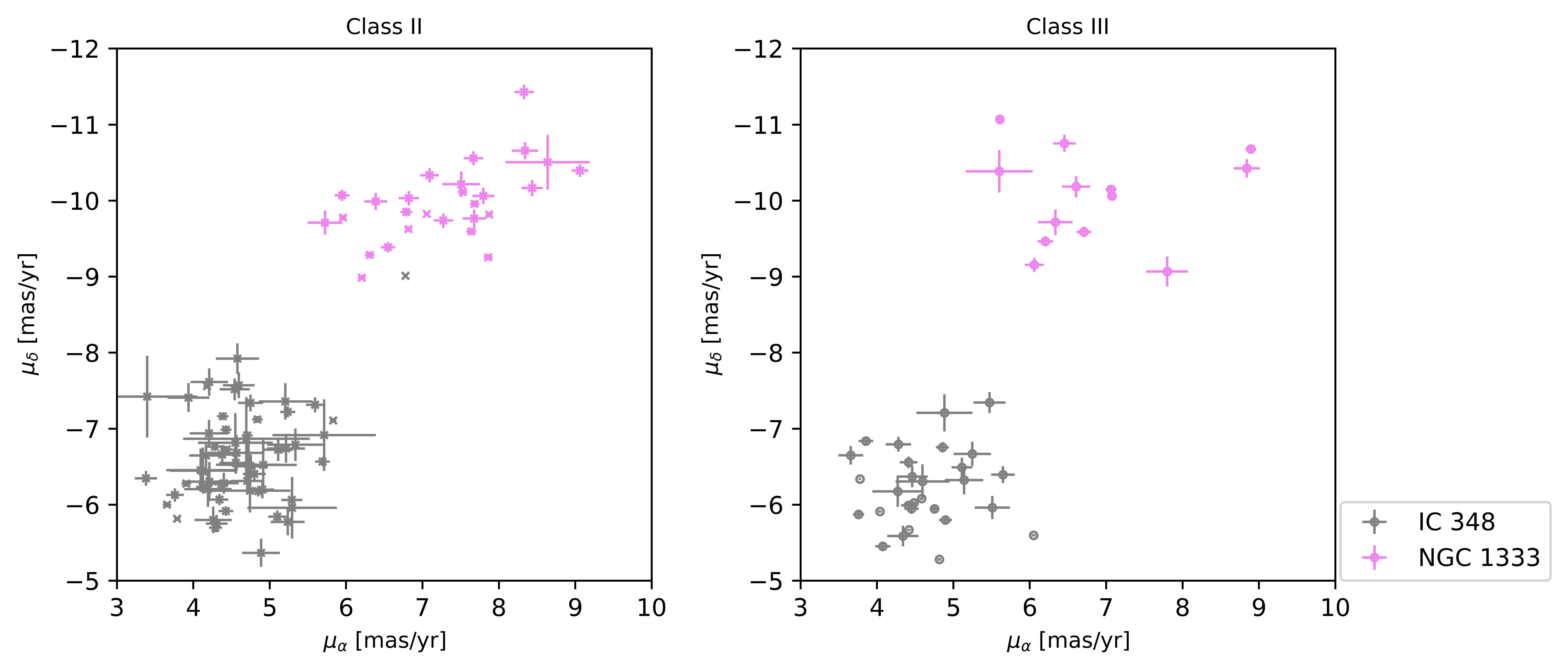}
	\caption{Proper motion distribution of YSOs in Perseus, where the crosses represent Class II YSOs and the circles represent Class III YSOs.}
	\label{fig:perseuspm}
\end{figure}

The projected distribution of YSOs in Perseus is shown in Figure \ref{fig:perseuspicyso}. The YSOs are mainly distributed along the dense regions of the molecular cloud and mainly focused in IC~348 and NGC~1333. In IC~348, the Class III YSOs are mainly concentrated towards the center, while the Class II YSOs are widely dispersed across the area. In contrast, both the Class II and III YSOs in NGC~1333 are compactly distributed in a small region with a scale of about 1 pc. Figure \ref{fig:perseuspm} displays the proper motion distributions of the YSOs in IC~348 and NGC~1333. The kinematic characteristics of these two regions show a significant difference in proper motion, concentrated at $(\mu_{\alpha^*}, \mu_{\delta})$ [4.62, -6.42] mas yr$^{-1}$ and [6.53, 10.12] mas yr$^{-1}$, respectively.

Table \ref{tab:Perseus info} lists the kinematic information of the YSOs in IC~348 and NGC~1333. The Class III YSOs in IC~348 show larger velocity offsets from the parent cloud in the line-of-sight direction compared with the Class II YSOs, where $|\Delta V_{\rm{LSR,rad}}|$ is 0.1 km s$^{-1}$ and 1.1 km s$^{-1}$ for the Class II and III YSOs, respectively. However, the situation in NGC~1333 is reversed, where the Class II YSOs show larger $|\Delta V_{\rm{LSR,rad}}|$ values than the Class III YSOs, with values of 0.9 and 0.5 km s$^{-1}$, respectively. $|\Delta V_{\rm{1D}}|$ is 1 km s$^{-1}$ in IC~348 and 0.4 km s$^{-1}$ in NGC~1333. The Class II YSOs in both IC~348 and NGC~1333 show smaller 3D velocities compared with the Class III YSOs, with $|\Delta V_{\rm{3D}}|$ values of 1.1 km s$^{-1}$ and 0.7 km s$^{-1}$, respectively. Additionally, the velocity dispersions of both $V_{\rm{LSR,rad}}$ and $V_{\rm{3D}}$ for the Class II YSOs are smaller than those for the Class III YSOs in both subregions.

\subsection{Taurus}

\begin{figure}
	\centering
	\includegraphics[scale=0.7]{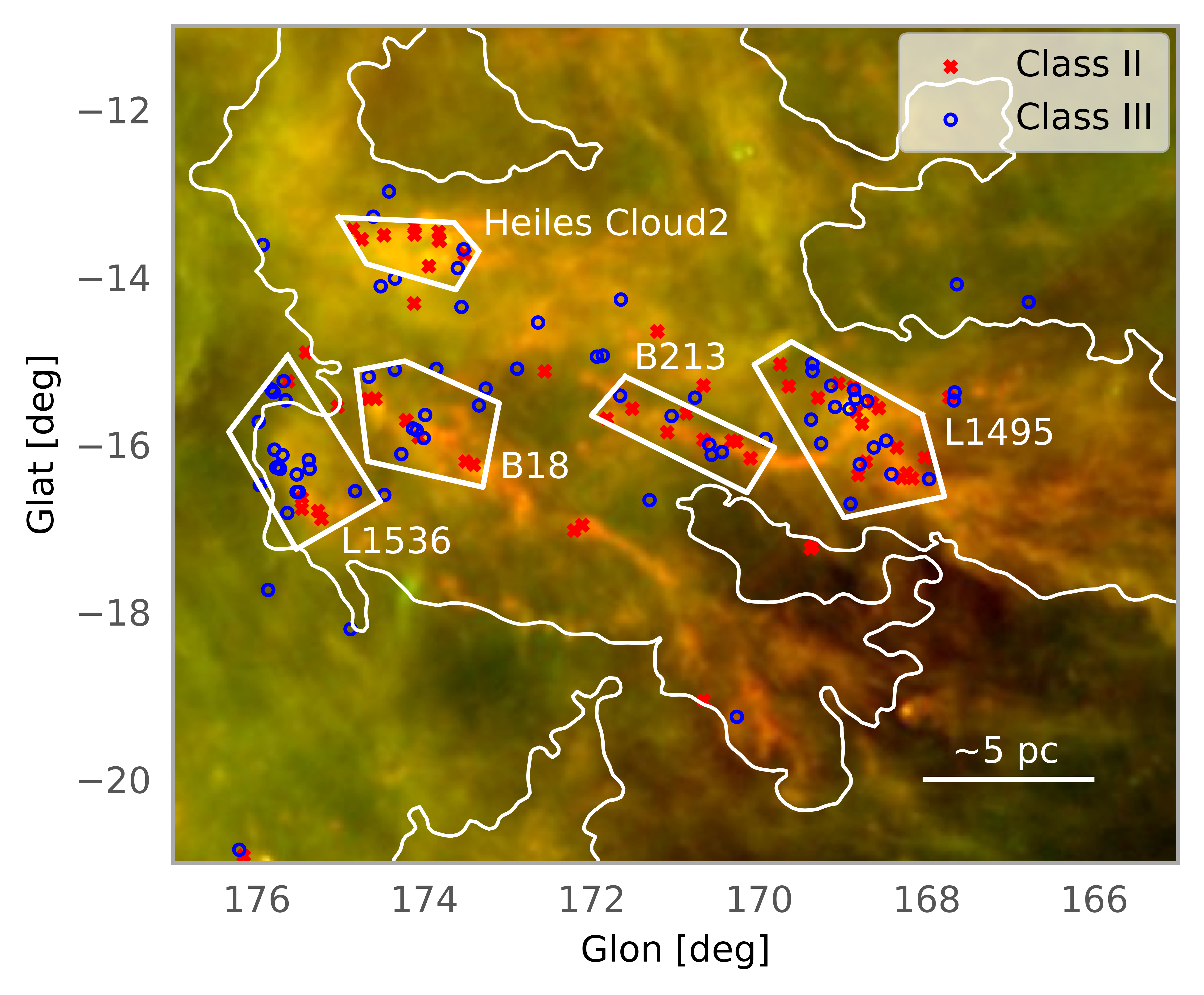}
	\caption{Projected distribution of YSOs in Taurus. The background is a false-color image of Taurus (red: Planck at 353 GHz; green: IRAS at 100 $\mu$m). The contours represent the velocity channel of [$-$2.6,15.6] km s$^{-1}$. The other features are the same as in Figure \ref{fig:orionapicyso}.}
	\label{fig:tauruspicyso}
\end{figure}

The Taurus molecular cloud covers about 100 deg$^2$ in the sky plane at a distance of $\sim$140 pc \citep{Galli2018}. It has a relatively low stellar density of around 1--10 pc$^{-3}$ \citep{Luhman2018AJ....156..271L} and massive stars do not appear to form within it. Young stars are roughly distributed along the filamentary structures of Taurus \citep{Hartmann2002}, accompanied by numerous molecular outflows and bubbles driven by stellar winds \citep[e.g.][]{Narayanan2012MNRAS.425.2641N, Li2015ApJS..219...20L}. There are approximately a dozen small star clusters associated with Taurus, with a total of more than 500 members \citep{Esplin2019}. These star clusters exhibit slightly different average velocities \citep{Luhman2023}. Following the results of \citet{Goldsmith2008ApJ...680..428G}, we split the Taurus molecular cloud into five subregions: Heiles' Cloud 2, L1495, L1536, B18, and B213, as shown in Figure \ref{fig:tauruspicyso}.

\begin{table*}
	\centering
	\caption{Kinematic Information of Young Stellar Objects in Taurus}
	\label{tab:Taurus info}
	\begin{tabular}{cccccccccccc}
		\hline
		\multirow{2}{*}{region} & \multirow{2}{*}{Class} & \multirow{2}{*}{N} & $\mu_{\alpha^*}$ & $\mu_\delta$ & $V_{\rm{LSR,rad}}$ & $\sigma_{V_{\rm{LSR,rad}}}$ & $|\Delta V_{\rm{LSR,rad}}|$&$|\Delta V_{\rm{1D}}|$ & $V_{\rm{3D}}$ & $\sigma_{V_{\rm{3D}}}$ & $|\Delta V_{\rm{3D}}|$ \\
		& & & mas yr$^{-1}$ & mas yr$^{-1}$ & km s$^{-1}$ & km s$^{-1}$ & km s$^{-1}$& km s$^{-1}$ & km s$^{-1}$ & km s$^{-1}$ & km s$^{-1}$ \\ \hline
		\multirow{4}{*}{Heiles' Cloud 2} & Cloud & - & - & - & 6.0 & 0.5 & - & - & - &- & - \\ 
		\cline{2-12} & I & 0 & - & - & - & - & - & - &- & - & - \\ 
		\cline{2-12} & II & 10 & 5.36$\pm$0.09 & -20.73$\pm$0.09 & 6.3$\pm$0.3 & 0.9 & 0.3$\pm$0.3 &- & 6.2$\pm$0.3 & 0.9 & - \\ 
		\cline{2-12} & III & 3 & 6.31$\pm$0.61 & -23.75$\pm$0.91 & 10.7$\pm$1.9 & 3.3 & 4.7$\pm$1.9 &4.4$\pm$2.0 & 10.9$\pm$1.7 & 2.9 & 4.7$\pm$1.7 \\ \hline
		\multirow{4}{*}{L1495} & Cloud & - & - & - & 6.8 & 1.0 & - & - & - &- & - \\ 
		\cline{2-12} & I & 1 & 9.70$\pm$0.4 & -25.93$\pm$0.22 & 7.8$\pm$0.2 & - & 1.0$\pm$0.1 &-& 9.0$\pm$0.5  &- & -\\ 
		\cline{2-12} & II & 15 & 8.60$\pm$0.06 & -25.01$\pm$0.06 & 6.8$\pm$0.4 & 1.5 & 0.0$\pm$0.4 &- & 7.1$\pm$0.3 & 1.2 & - \\ 
		\cline{2-12} & III & 12 & 8.75$\pm$0.07 & -25.61$\pm$0.07 & 6.4$\pm$1.1 & 3.7 & 0.4$\pm$1.1 &0.4$\pm$1.2 & 7.0$\pm$0.9 & 3.1 & 0.1$\pm$0.9 \\ \hline
		\multirow{4}{*}{L1536} & Cloud & - & - & - & 5.7 & 0.8 & - & - & - &- & - \\ 
		\cline{2-12} & I & 0 & - & - & - & - & - & - & - &- & - \\ 
		\cline{2-12} & II & 4 & 8.95$\pm$0.24 & -16.64$\pm$0.14 & 5.9$\pm$0.5 & 1.0 & 0.2$\pm$0.5 &- & 6.5$\pm$0.2 & 0.5 & - \\ 
		\cline{2-12} & III & 11 & 9.78$\pm$0.13 & -17.55$\pm$0.32 & 7.4$\pm$0.9 & 3.1 & 1.7$\pm$0.9 &1.5$\pm$1.0 & 8.9$\pm$0.7 & 2.3 & 2.4$\pm$0.7 \\ \hline
		\multirow{4}{*}{B18} & Cloud & - & - & - & 6.0 & 0.4 & - & - &- & - & - \\ 
		\cline{2-12} & I & 0 & - & - & - & - & - & - &- & - & - \\ 
		\cline{2-12} & II & 8 & 7.17$\pm$0.14 & -21.03$\pm$0.05 & 6.7$\pm$0.6 & 1.6 & 0.7$\pm$0.6 &- & 6.4$\pm$0.5 & 1.5 & - \\ 
		\cline{2-12} & III & 11 & 6.99$\pm$0.24 & -21.57$\pm$0.19 & 5.7$\pm$0.5 & 1.6 & 0.3$\pm$0.5 &1.0$\pm$0.8 & 5.8$\pm$0.5 & 1.7 & 0.6$\pm$0.7 \\ \hline
		\multirow{4}{*}{B213} & Cloud & - & - & - & 6.4 & 0.5 & - & - &- & - & - \\ 
		\cline{2-12} & I & 0 & - & - & - & - & - & - & - & - &- \\ 
		\cline{2-12} & II & 6 & 11.42$\pm$0.03 & -17.69$\pm$0.07 & 7.0$\pm$0.4 & 0.9 & 0.6$\pm$0.4 &- & 7.9$\pm$0.3 & 0.7 & - \\ 
		\cline{2-12} & III & 5 & 11.60$\pm$0.06 & -17.49$\pm$0.07 & 5.3$\pm$1.6 & 3.5 & 1.1$\pm$1.6 &1.7$\pm$ 1.6& 7.3$\pm$0.9 & 2.0 & 0.6$\pm$0.9 \\ \hline
	\end{tabular}
\end{table*}

The star clusters in the five subregions exhibit different parallaxes and proper motions \citep{Luhman2018AJ....156..271L}. Therefore, we applied different parallax criteria to the YSOs in each subregion (see Table \ref{tab:parallax limit} and Appendix \ref{sec:appendix}). In total, we selected 85 YSOs in Taurus with precise distances and 3D kinematics, including 43 Class II and 42 Class III YSOs. The number of YSOs in each subregion is listed in Table \ref{tab:Taurus info}.

\begin{figure}
	\centering
	\includegraphics[scale=0.6]{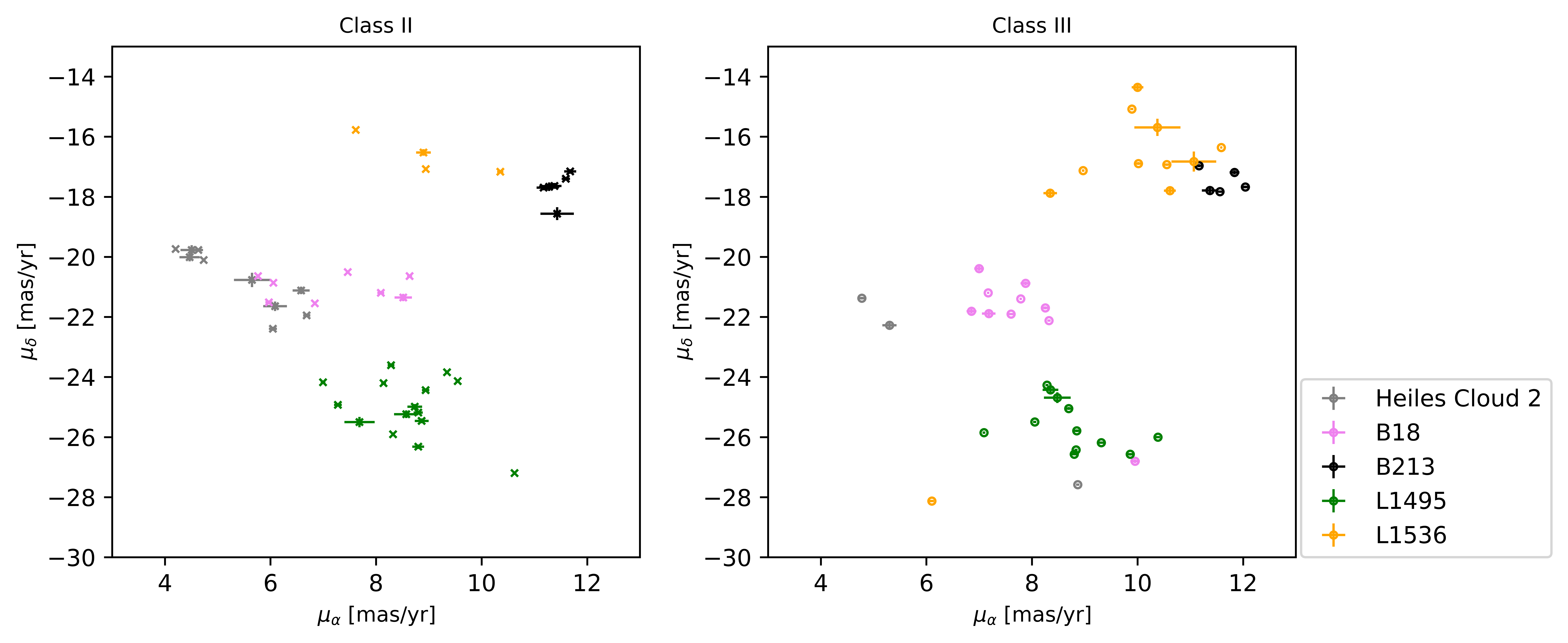}
	\caption{Proper motion distribution of YSOs in Taurus, where the crosses represent Class II YSOs and the circles represent Class III YSOs.}
	\label{fig:tauruspm}
\end{figure}

Figure \ref{fig:tauruspicyso} shows the projected distribution of the YSOs in Taurus. The YSOs in the five subregions are mainly located in the densest parts of the molecular cloud. There is a long filamentary structure extending from B213 to L1495, and the YSOs in these two subregions show different parallaxes, i.e., 6.22$\pm$0.03 mas (160$\pm$0.9 pc) and 7.66$\pm$0.02 mas (130$\pm$0.4 pc), respectively. Despite the YSOs in B18 and L1536 being located close in the sky plane, their parallaxes show significant differences \citep{Luhman2018AJ....156..271L}. The average parallax of the YSOs in B18 is 7.78$\pm$0.01 mas (129$\pm$0.2 pc), while in L1536 the value is 6.17$\pm$0.04 mas (162$\pm$1 pc). The average parallax of the YSOs in Heiles' Cloud 2 is about 7.13$\pm$0.04 mas (140$\pm$0.8 pc). Figure \ref{fig:tauruspm} shows the proper motion distribution of the YSOs in Taurus. YSOs in the same subregion exhibit similar kinematic characteristics, whereas the characteristics differ among the different subregions.

Table \ref{tab:Taurus info} presents the LSR velocity offsets from the parent molecular cloud and the 3D velocities of the YSOs. We only analyzed the kinematic characteristics in four subregions, Heiles' Cloud 2, L1495, L1536, and B18, since only these regions contain more than 10 Class II or III YSOs. In L1495, the Class II YSOs show no offset from the parent cloud (their $|\Delta V_{\rm{LSR,rad}}|$ value is $\sim$ 0 km s$^{-1}$). The Class III YSOs in this region show a small offset of 0.4 km s$^{-1}$. The $|\Delta V_{\rm{1D}}|$ is 0.4 km s$^{-1}$ in L1495. Meanwhile, the Class II and III YSOs show similar 3D kinematic characteristics in this region, with $|\Delta V_{\rm{3D}}|$ being only 0.1 km s$^{-1}$, while the Class II YSOs exhibit a smaller velocity dispersion than the Class III YSOs. The YSOs in B18 exhibit kinematic characteristics similar to their parent molecular cloud, with a $|\Delta V_{\rm{LSR,rad}}|$ of 0.3 km s$^{-1}$. However, the Class III YSOs in L1536 show a larger velocity offset (1.7 km s$^{-1}$) from their parent cloud in the line-of-sight direction compared to the other subregions.

\subsection{$\lambda$~Orionis}

$\lambda$~Orionis is a highly evolved star-forming region, containing stars with masses ranging from 0.2 M${_\odot}$ to 24 M${_\odot}$. The shape of $\lambda$~Orionis is characterized by a ring of neutral and molecular gas. According to the research of \citet{Dolan2002AJ....123..387D}, the ring-like structure in this region was mainly produced by the supernova explosion of O-type binary stars that occurred approximately 1 Myr ago. The remaining massive O8-type star $\lambda$~Ori formed the H~\textsc{ii} region. A massive cluster (Collinder~69), containing $\lambda$~Ori, several B-type stars, and numerous low-mass stars and brown dwarfs \citep{Bayo2011}, is situated at the center, while the dark clouds B30, B35, and B223 are located on the edge of the H~\textsc{ii} region, where star formation is still ongoing.  

\begin{table*}
	\centering
	\caption{Kinematic Information of the Young Stellar Objects in $\lambda$~Orionis}
	\label{tab:Orionis info}
	\begin{tabular}{cccccccccccc}
		\hline
		\multirow{2}{*}{region} & \multirow{2}{*}{Class} & \multirow{2}{*}{N} & $\mu_{\alpha^*}$ & $\mu_\delta$ & $V_{\rm{LSR,rad}}$ & $\sigma_{V_{\rm{LSR,rad}}}$ & $|\Delta V_{\rm{LSR,rad}}|$ &$|\Delta V_{\rm{1D}}|$& $V_{\rm{3D}}$ & $\sigma_{V_{\rm{3D}}}$ & $|\Delta V_{\rm{3D}}|$ \\
		& & & mas yr$^{-1}$ & mas yr$^{-1}$ & km s$^{-1}$ & km s$^{-1}$ & km s$^{-1}$ & km s$^{-1}$& km s$^{-1}$ & km s$^{-1}$ & km s$^{-1}$ \\ \hline
		\multirow{4}{*}{B30} & Cloud & - & - & - & 10.2 & 1.1 & - &- & - & - & - \\ 
		\cline{2-12} & I & 0 & - & - & - & - & - & - &- & - & - \\ 
		\cline{2-12} & II & 20 & 0.91$\pm$0.02 & -0.76$\pm$0.04 & 11.2$\pm$0.1 & 2.2 & 1.0$\pm$0.4 &- & 12.6$\pm$0.1 & 1.8 & - \\ 
		\cline{2-12} & III & 10 & 0.87$\pm$0.08 & -0.91$\pm$0.06 & 11.3$\pm$0.2 & 1.7 & 1.1$\pm$0.4 &0.1$\pm$0.2 & 12.5$\pm$0.2 & 1.7 & 0.1$\pm$0.2 \\ \hline
	\end{tabular}
\end{table*}

\begin{figure}
	\centering
	\includegraphics[scale=0.6]{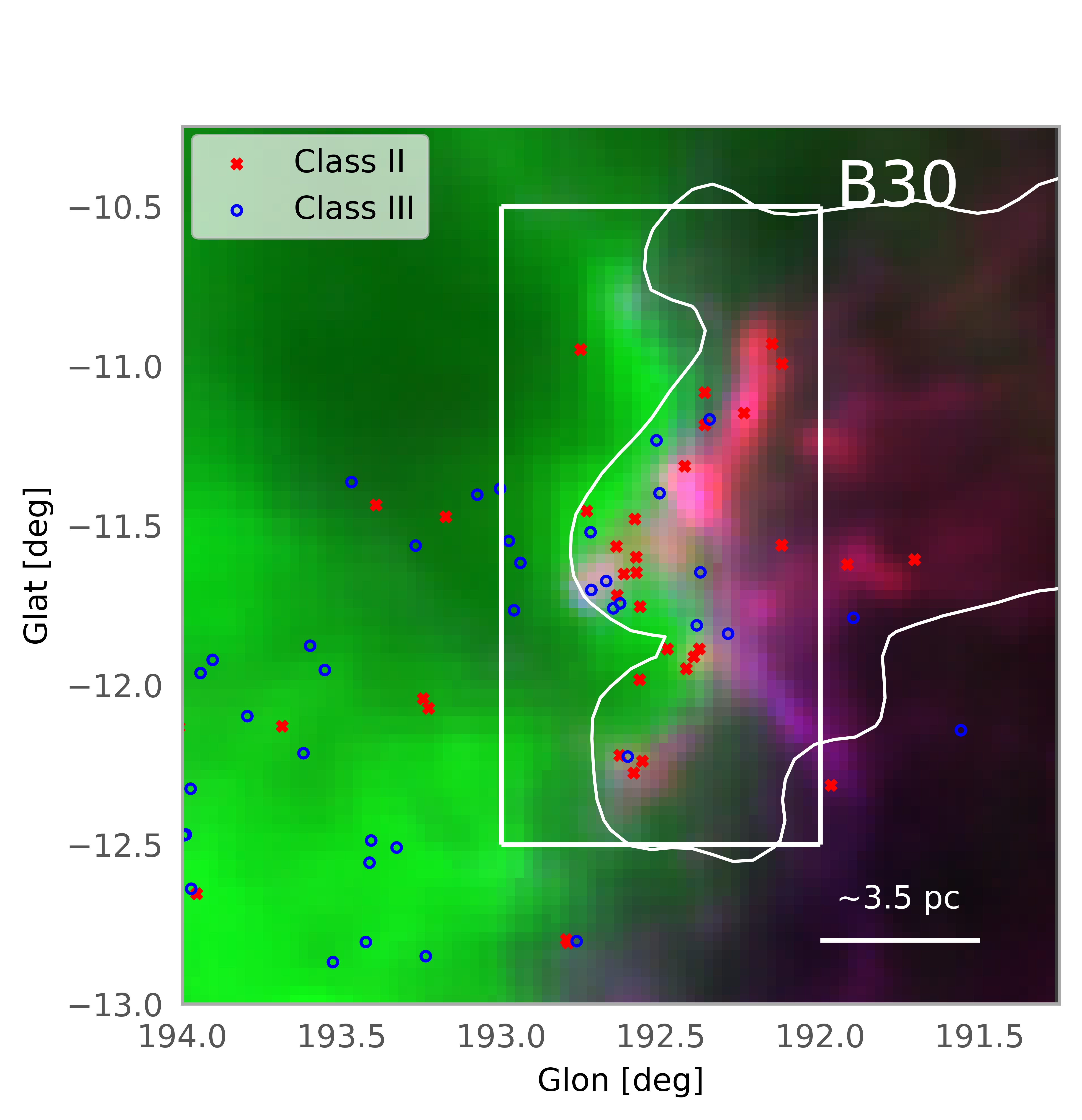}
	\caption{Projected distribution of YSOs in $\lambda$~Orionis. The background is a false-color image of $\lambda$~Orionis (red: Planck 353 GHz; blue: IRAS 100 $\mu$m; and green: H$\alpha$). The contours represent the velocity channel of [-2.6, 15.6] km s$^{-1}$. The other features are the same as in Figure \ref{fig:orionapicyso}.   }
	\label{fig:lamOripicyso}
\end{figure}

In this work, we focus on B30 to investigate the relationship between the gas and stars, as this region is influenced by massive stars and supernova feedback. The detailed parallax criteria are outlined in Appendix \ref{sec:appendix} and Table \ref{tab:parallax limit}. In all, 30 YSOs in B30 with precise distance and 3D kinematics measurements were selected, including 20 Class II and 10 Class III YSOs (see details in Table \ref{tab:Orionis info}). Figure \ref{fig:lamOripicyso} displays the projected distribution of the YSOs in $\lambda$~Orionis. The YSOs in B30 are primarily distributed at the interface between the molecular cloud and the H~\textsc{ii} region, which may be due to the latter occupying most of the star-forming region and continuously compressing and disrupting the surrounding molecular clouds \citep{Yi2018}. Meanwhile, the Class II YSOs in B30 show greater aggregation than their Class III counterparts. The proportions of YSOs in the bubble and B30 are 67\% and 30\% for the Class III YSOs, respectively, indicating that the star formation activity in B30 is not as advanced as in the bubble. 

\begin{figure}
	\centering
	\includegraphics[scale=0.6]{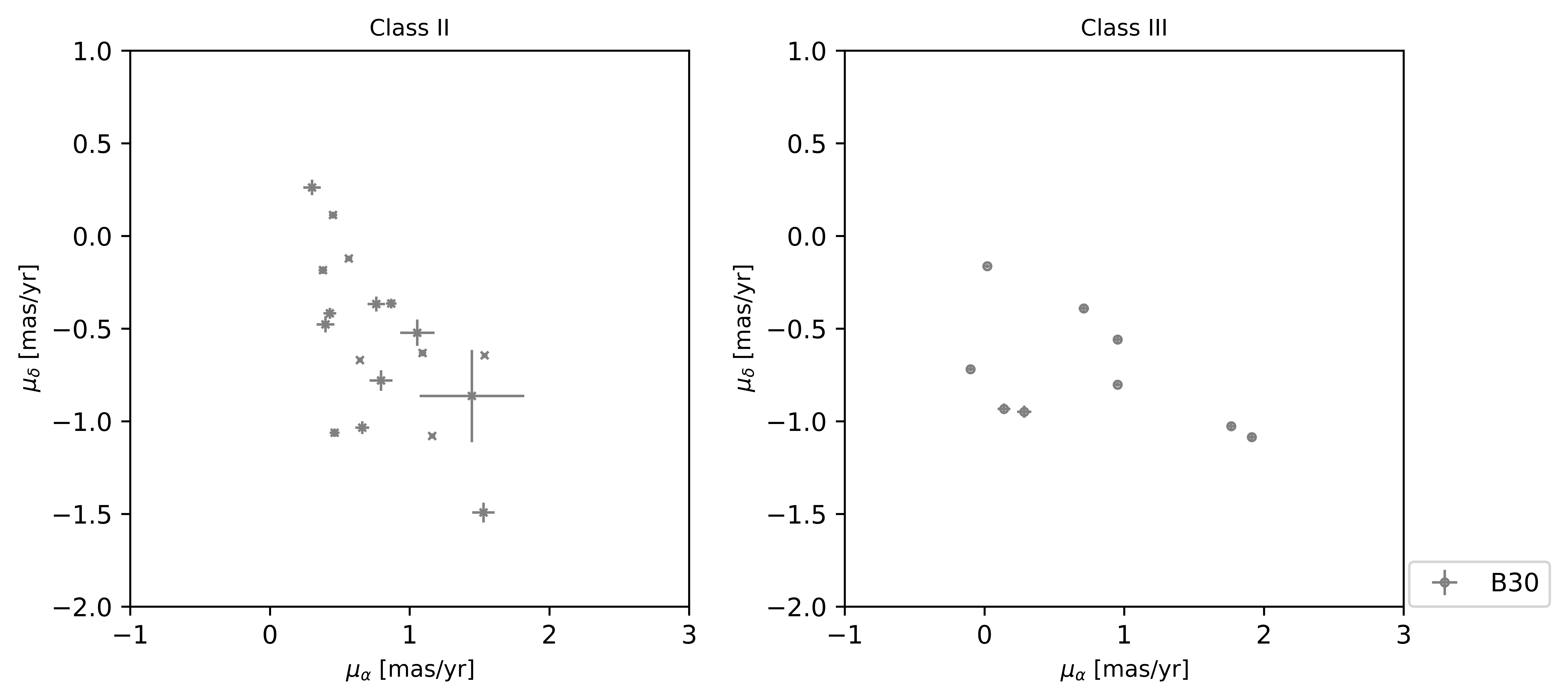}
	\caption{Proper motion distribution of YSOs in B30. The crosses represent Class II YSOs and the circles represent Class III YSOs.}
	\label{fig:b30pm}
\end{figure}

Figure \ref{fig:b30pm} show the proper motion distribution of the YSOs in B30 and Table \ref{tab:Orionis info} lists their kinematic information, including their offsets from the parent cloud and average 3D velocity. The Class II and III YSOs in B30 show similar kinematic characteristics, including motion along the celestial plane, line-of-sight direction, and 3D velocity. Therefore, the Class II and III YSOs exhibit similar velocity offsets from the parent cloud, i.e., with $|\Delta V_{\rm{LSR,rad}}|$ values of 1.0 km s$^{-1}$ and 1.1 km s$^{-1}$, respectively. Both $|\Delta V_{\rm{1D}}|$ and $|\Delta V_{\rm{3D}}|$ are 0.1 km s$^{-1}$.  

\section{Kinematic Characteristics of the Young Stellar Objects}
\label{discussion}

\begin{figure}[htbp]
	\centering
	\subfigure[a]{\includegraphics[scale=0.5]{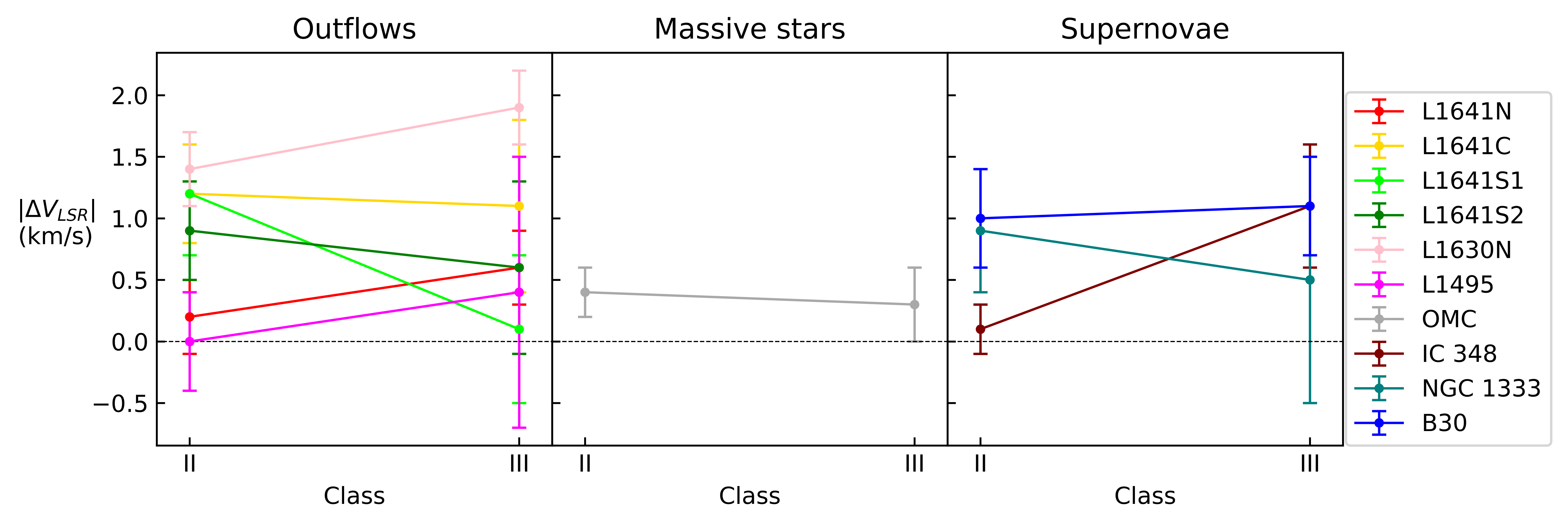}\label{vlsr}}
	\subfigure[b]{\includegraphics[scale=0.5]{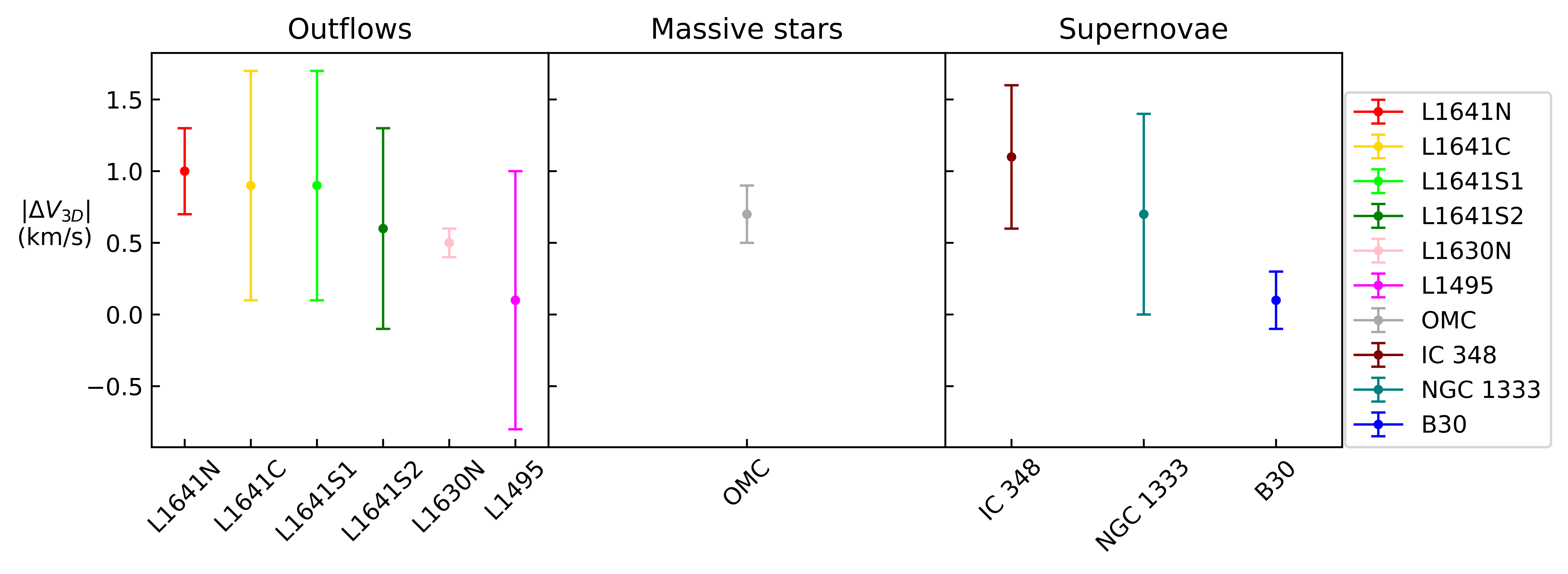}\label{v3d}}
	\caption{Kinematics characteristics of the studied star-forming regions. (a) The difference in the LSR velocity between different classes of YSOs and molecular gas, i.e., $|\Delta V_{\rm{LSR,rad}}|$. (b) The difference in 3D velocity between the Class II and III YSOs, i.e., $|\Delta V_{\rm{3D}}|$. From left to right the feedback environment consists of outflows, massive stars, and supernovae. The points in different colors represent the investigated subregions.}
	\label{fig:vlsrandv3d}
\end{figure}

\subsection{Kinematic Differences among the Different Evolutionary Stages of Young Stellar Objects}

Due to the limited number of YSOs in some subregions, we only analyzed the kinematic characteristics in those with more than 10 Class II or III YSOs. These subregions are as follows: (1) both the Class II and III YSOs exceed 10 samples, i.e, the OMC, L1641N, L1641C, L1641S1, and L1641S2 in Orion~A, L1630N in Orion~B, IC~348 and NGC~1333 in Perseus, L1495 in Taurus, and B30 in $\lambda$~Orionis; (2) only the Class II YSOs exceed 10 samples, i.e, NGC~2024 in Orion~B and Heiles' Cloud 2 in Taurus; and (3) only the Class III YSOs exceed 10 samples, i.e, B18 and L1536 in Taurus. 

Figure \ref{fig:vlsrandv3d} shows the $|\Delta V_{\rm{LSR,rad}}|$ and $|\Delta V_{\rm{3D}}|$ results. The $|\Delta V_{\rm{LSR,rad}}|$ values of the Class III YSOs are smaller than their Class II counterparts in five subregions, i.e., OMC, L1641C, L1641S1, L1641S2 and NGC 1333. Both the Class II and III YSOs show small offsets from their parent clouds, where the highest $|\Delta V_{\rm{LSR,rad}}|$ value is 1.4 km s$^{-1}$ and 1.9 km s$^{-1}$, respectively. The maximum $|\Delta V_{\rm{3D}}|$ value is 1.1 km s$^{-1}$. 

Furthermore, except for L1495 and NGC~1333, the Class III YSOs in the other subregions exhibit larger LSR velocities compared with the Class II YSOs therein. Except for L1630N and B30, the Class III YSOs in the other subregions exhibit larger LSR velocity dispersions compared with the Class II YSOs. The mean LSR velocity dispersions for both the Class II and III YSOs are similar, i.e., 2.1 km s$^{-1}$ and 2.7 km s$^{-1}$, respectively. The $V_{\rm{3D}}$ of the Class III YSOs in L1495 and B30 is smaller than for the Class II YSOs in the same region. The average dispersion of $V_{\rm{3D}}$ for the Class II YSOs is 1.8 km s$^{-1}$, while it is 2.4 km s$^{-1}$ for the Class III YSOs. 

These results indicate that the Class II YSOs are better correlated with their parent molecular clouds, whereas the 3D motions of the Class III YSOs larger and more random.

\subsection{Impact of Feedback Environment on Young Stellar Object Kinematics}

Feedback can significantly influence the kinematic evolution of YSOs when they emerge from their parent molecular clouds. The five clouds we selected exhibit diverse feedback environments, including outflows, photoionization, radiation pressure and thermal stellar winds in H~\textsc{ii} regions, and supernova explosions. Investigating the kinematics of YSOs under the effects of these different types of stellar feedback can be achieved by calculating the maximum velocity of YSOs emerging from their parent molecular clouds.

Based on the different feedback environments, we classified the subregions into three categories: (1) outflows, including L1641N, L1641C, L1641S1, L1641S2, L1630N, Heiles' Cloud 2 (only Class II YSOs are counted), L1495, L1536, and B18 (only Class III YSOs are counted); (2) massive stars, including the OMC and NGC~2024 (only Class II YSOs are counted); (3) supernova explosions, including IC~348, NGC~1333 and B30.

In clouds with outflows, the Class II YSOs in only three subregions (i.e., L1641C, L1641S1, and L1630N) exhibit an LSR velocity offset ($|\Delta V_{\rm{LSR,rad}}|$) more than 1 km s$^{-1}$ from their parent cloud, with the maximum offset being 1.4 km s$^{-1}$ in L1630N. Similarly, the $|\Delta V_{\rm{LSR,rad}}|$ values of the Class III YSOs exceed 1 km s$^{-1}$ in L1630N, L1536, and L1641C. The $|\Delta V_{\rm{LSR,rad}}|$ value for the Class III YSOs in L1641C is 1.1 km s$^{-1}$. Although the values for L1630N and L1536 are 1.9 km s$^{-1}$ and 1.7 km s$^{-1}$, respectively, these may not necessarily be reliable, since only 13 and 10 Class III YSOs were selected in L1630N and L1536, respectively. The maximum $|\Delta V_{\rm{1D}}|$ is 1.1 km s$^{-1}$ in L1641S1. The maximum $|\Delta V_{\rm{3D}}|$ value is 1 km s$^{-1}$ for the YSOs in L1641N. This might be due to molecular cloud collisions occurring in L1641N, as suggested by \citet{Nakamura2012}.

For star-forming regions affected by feedback from massive stars, the LSR velocity offsets between all YSOs and their parent clouds are relatively small, with the maximum offset being 0.4 km s$^{-1}$. Meanwhile, both the maximum $|\Delta V_{\rm{1D}}|$ and $|\Delta V_{\rm{3D}}|$ are 0.7 km s$^{-1}$. Note that, because there are only four Class III YSOs in NGC~2024, we did not analyze the kinematics of the Class III YSOs for this subregion. All YSOs in the OMC and NGC~2024 exhibit small $|\Delta V_{\rm{LSR,rad}}|$ and $|\Delta V_{\rm{3D}}|$ values. Therefore, feedback from massive stars may not significantly impact the kinematics of the YSOs during the process of emerging from their parent molecular cloud.

\begin{figure}[htbp]
	\centering
	\subfigure[OMC]{\includegraphics[scale=0.5]{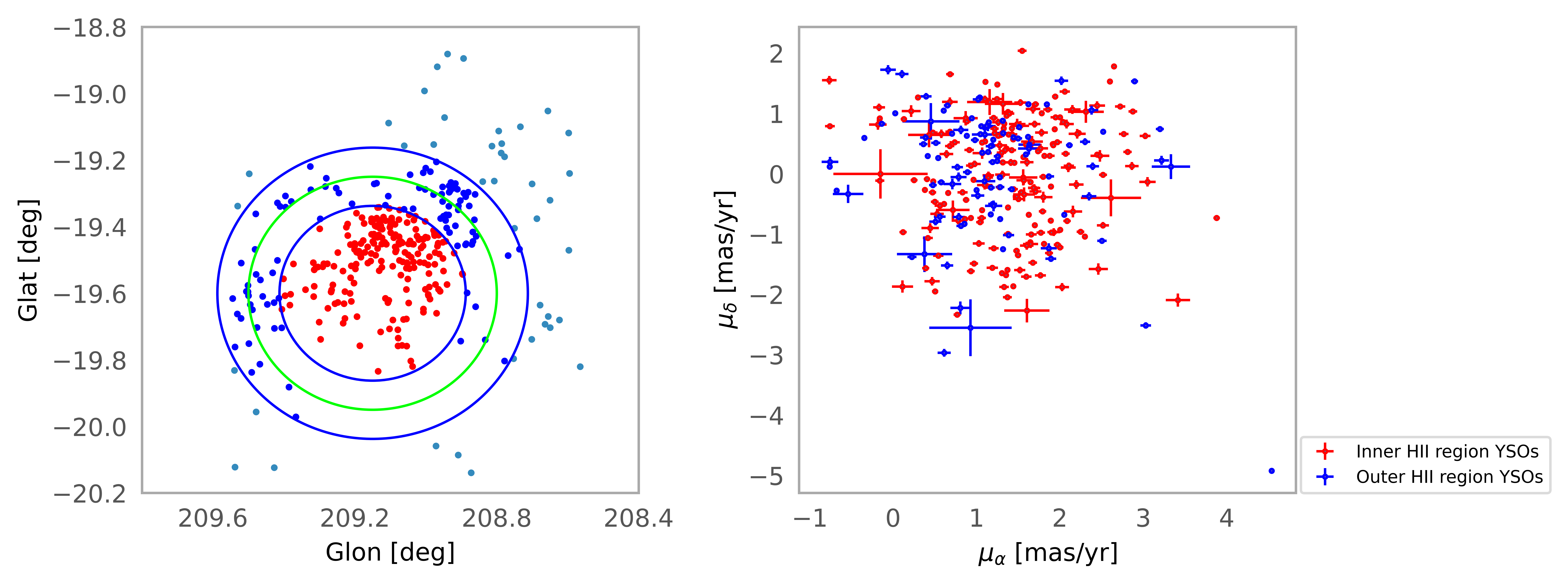}\label{OMC_h2}}
	\subfigure[NGC 2024]{\includegraphics[scale=0.5]{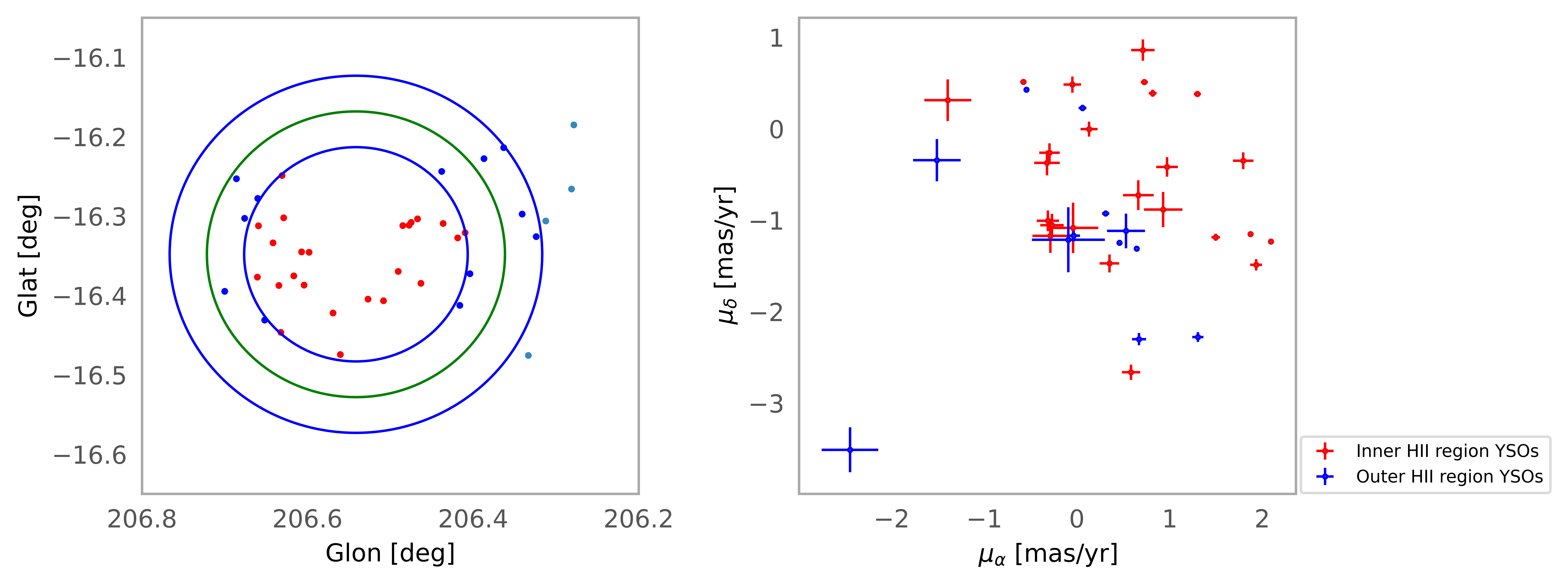}\label{NGC2024_h2}}
	\caption{This figure shows the kinematic characteristics of the star-forming regions affected by feedback from massive stars. (a) YSOs distribution in OMC. The green circle represents H~\textsc{ii} region the same as in \ref{fig:orionapicyso}. The blue circles represent regions with radii of three-fourths and five-fourths of H~\textsc{ii} region radii, respectively. The red dots represent YSOs inside H~\textsc{ii} region and the blue dots represent YSOs near the edge of H~\textsc{ii} region. (b) YSOs distribution in NGC 2024. The green circle represents H~\textsc{ii} region the same as in Figure \ref{fig:orionbpicyso}. The blue circles represent regions with radii of three-fourths and five-fourths of H~\textsc{ii} region radii, respectively. The red dots represent YSOs inside H~\textsc{ii} region and the blue dots represent YSOs near the edge of H~\textsc{ii} region and the blue dots represent YSOs near the edge of H~\textsc{ii} region.}
	\label{fig:HIIcomp}
\end{figure}

There are no significant differences in the proper motion distribution between YSOs inside and near the edge of the H \textsc{ii} region. Figure \ref{fig:HIIcomp} shows the kinematic characteristics of the star-forming regions affected by feedback from massive stars. The average $V_{\rm{LSR,rad}}$ of YSOs inside and near the edge of the H \textsc{ii} region are 8.4$\pm$0.2 km s$^{-1}$ and 9.5$\pm$0.3 km s$^{-1}$ in OMC, 10.0$\pm$0.5 km s$^{-1}$ and 9.3$\pm$1.2 km s$^{-1}$ in NGC 2024, respectively. The average $V_{\rm{3D}}$ of YSOs inside and near the edge of the H \textsc{ii} region are 7.2$\pm$0.1 km s$^{-1}$ and 8.2$\pm$0.3 km s$^{-1}$ in OMC, 8.7$\pm$0.4 km s$^{-1}$ and 8.0$\pm$0.9 km s$^{-1}$ in NGC 2024, respectively. We have conducted Mann-Whitney U tests for both subregions to compare the values of $V_{\rm{LSR,rad}}$ and $V_{\rm{3D}}$ of YSOs inside and near the edge of H \textsc{ii} region are larger, performing all tests at a significance level of $\alpha$ = 0.05. Results indicate that in NGC 2024, both $V_{\rm{LSR,rad}}$ and $V_{\rm{3D}}$ of YSOs located within or near the edges of the H \textsc{ii} regions show no significant differences. However, in OMC, the $V_{\rm{LSR,rad}}$ and $V_{\rm{3D}}$ of YSOs located within the H \textsc{ii} regions are smaller than those near the edges.

There are three subregions affected by supernova explosions. The maximum $|\Delta V_{\rm{LSR,rad}}|$ value for the Class II YSOs is 1 km s$^{-1}$ for B30, while the maximum value for the Class III YSOs is 1.1 km s$^{-1}$ for IC~348 and B30. The maximum $|\Delta V_{\rm{1D}}|$ and $|\Delta V_{\rm{3D}}|$ values are for IC~348, i.e., 1 km s$^{-1}$ and 1.1 km s$^{-1}$, respectively. IC~348 shows significant kinematic differences between its Class II and III YSOs in terms of LSR and 3D velocities. 

These results indicate that the LSR velocity differences between the YSOs and their parent clouds are quite small, generally less than 2 km s$^{-1}$. The results are similar to the most probable velocity dispersion of 1.95 km s$^{-1}$ by \citet{Miville2017}, who investigate 8107 molecular clouds that covers the entire Galactic plane. We conducted Mann-Whitney U tests to determine whether the $|\Delta V_{\rm{LSR,rad}}|$ of Class II YSOs are greater than Class III YSOs across three categories, with all tests performed at a significance level of $\alpha$ = 0.05, taking errors into account. The results showed that in clouds with outflows, there are no significant $|\Delta V_{\rm{LSR,rad}}|$ differences between Class II and Class III YSOs. However, in the clouds affected by massive stars and supernova explosions, the $|\Delta V_{\rm{LSR,rad}}|$ of Class III YSOs is significantly greater than Class II YSOs, and this conclusion remained valid even after considering errors. The 3D velocity differences between the Class II and III YSOs are also small for the different types of feedback environments.

\section{Conclusion}
\label{conclusion}

Combining high-precision astrometric data from Gaia DR3, high-precision radial velocities from APOGEE, and large-scale CO survey data obtained with the CfA 1.2 m telescope, we studied the kinematics of five star-forming regions (i.e., Orion~A, Orion~B, Perseus, Taurus, and $\lambda$~Orionis) within 500 pc of the Sun. By comparing the 1D velocity differences between the Class II YSOs therein and their parent molecular clouds, we found that the impact of the feedback processes on the kinematics of the YSOs is minimal, generally less than 2 km s$^{-1}$. In environments dominated by outflows, massive stars, and supernova feedback, there were no significant differences in the momentum injection. These results provide a valuable reference for studying the kinematics of the star formation process.

In the five star-forming regions studied in this paper, the 1D velocity differences between the Class III YSOs and Class II YSOs span [0.1, 1.1] km s$^{-1}$. Additionally, in environments dominated by outflows, massive stars, and supernova feedback, the corresponding velocity differences between the Class II YSOs and their parent clouds are [0, 1.4], [0.1, 0.4], and [0.1, 1] km s$^{-1}$, respectively. The Class III YSOs generally exhibit larger velocities and greater velocity dispersions compared to the Class II YSOs. This is probably because the Class III YSOs emerge from their parental molecular clouds earlier than the Class II YSOs.

In the future, Gaia data will provide more and higher-quality astrometric data for YSOs. A large amount of YSO samples we collected lack Gaia's full astrometric solutions. With more YSO samples and more measurements, it would expand the sample size and make the results more accurate. Combined with higher-precision radial velocity survey data and molecular gas survey data, we anticipate conducting kinematic studies of the star formation process in more star-forming regions.

\begin{acknowledgements}
This work was funded by the NSFC grant 11933011, National SKA Program of China (grant No. 2022SKA0120103), and the Key Laboratory for Radio Astronomy. Y.J.L. thanks supports from the NSFC grant 12203104 and the Natural Science Foundation of Jiangsu Province (grant No. BK20210999). This work is supported by the National Natural Science Foundation of China (grant No. 12133003). This work is also supported by the Guangxi Talent Program ("Highland of Innovation Talents"). ZJL acknowledges support by the grant AD23026127 and 2024GXNSFBA010436, funded by Guangxi Science and Technology Project. This work has made use of data from the European Space Agency (ESA) mission Gaia (\url{https://www.cosmos.esa.int/gaia}), processed by the Gaia Data Processing and Analysis Con sortium (DPAC, \url{https://www.cosmos.esa.int/web/gaia/dpac/consortium}). Funding for the DPAC
has been provided by national institutions, in particular the institutions participating in the Gaia Multilateral Agreement. This work has also made use of the VizieR, SIMBAD, and Aladin databases operated at CDS, Strasbourg, France.
\end{acknowledgements}
\appendix

\section{Parallax Distribution and Criteria Applied to the Young Stellar Object Samples}
\label{sec:appendix}

\setcounter{figure}{0} 
\renewcommand\thefigure{A.\arabic{figure}} 

\begin{figure}[htbp]
	\centering
	\subfigure[OMC]{\includegraphics[scale=0.6]{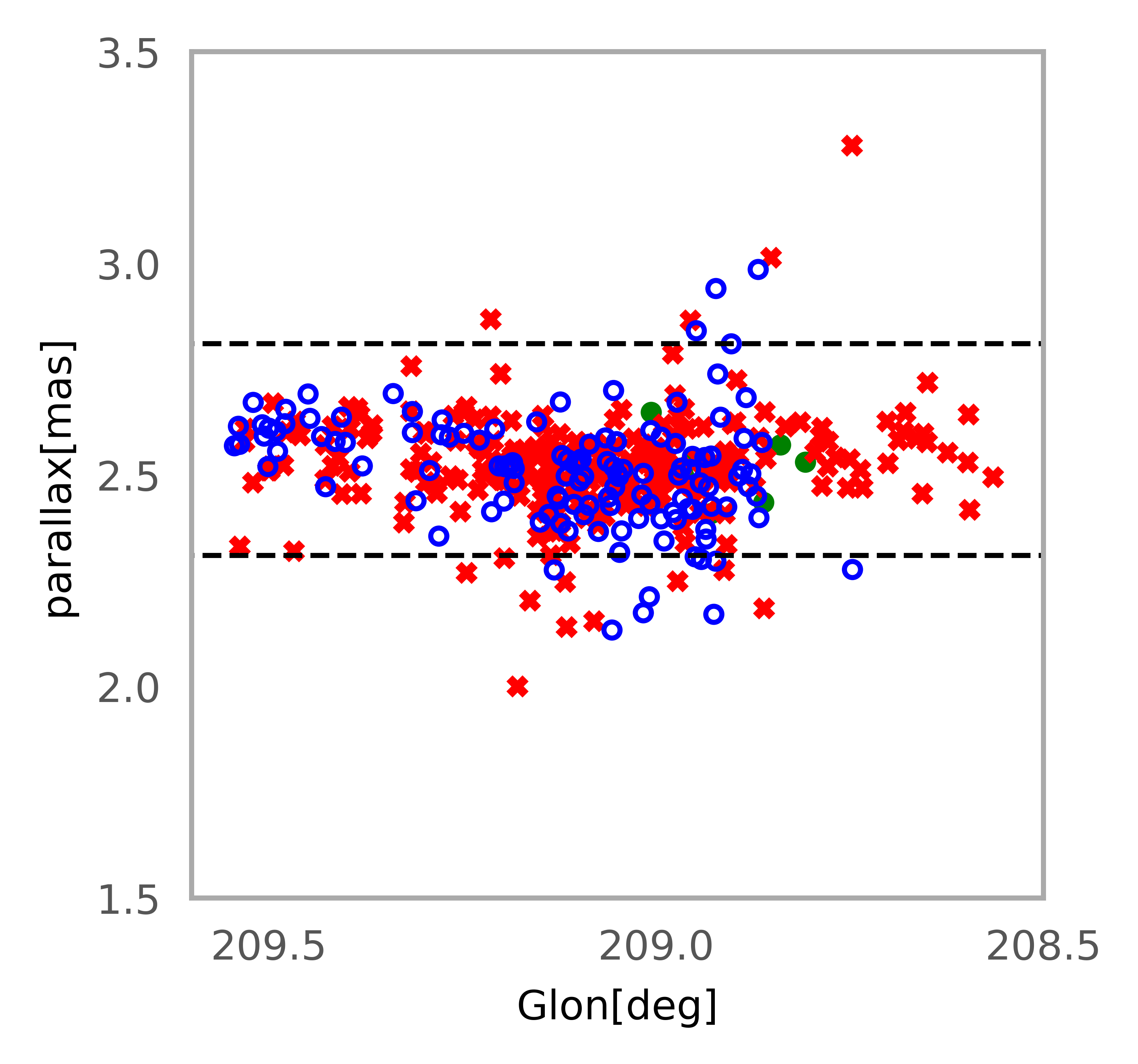}\label{omcp}}
	\subfigure[L1641N]{\includegraphics[scale=0.6]{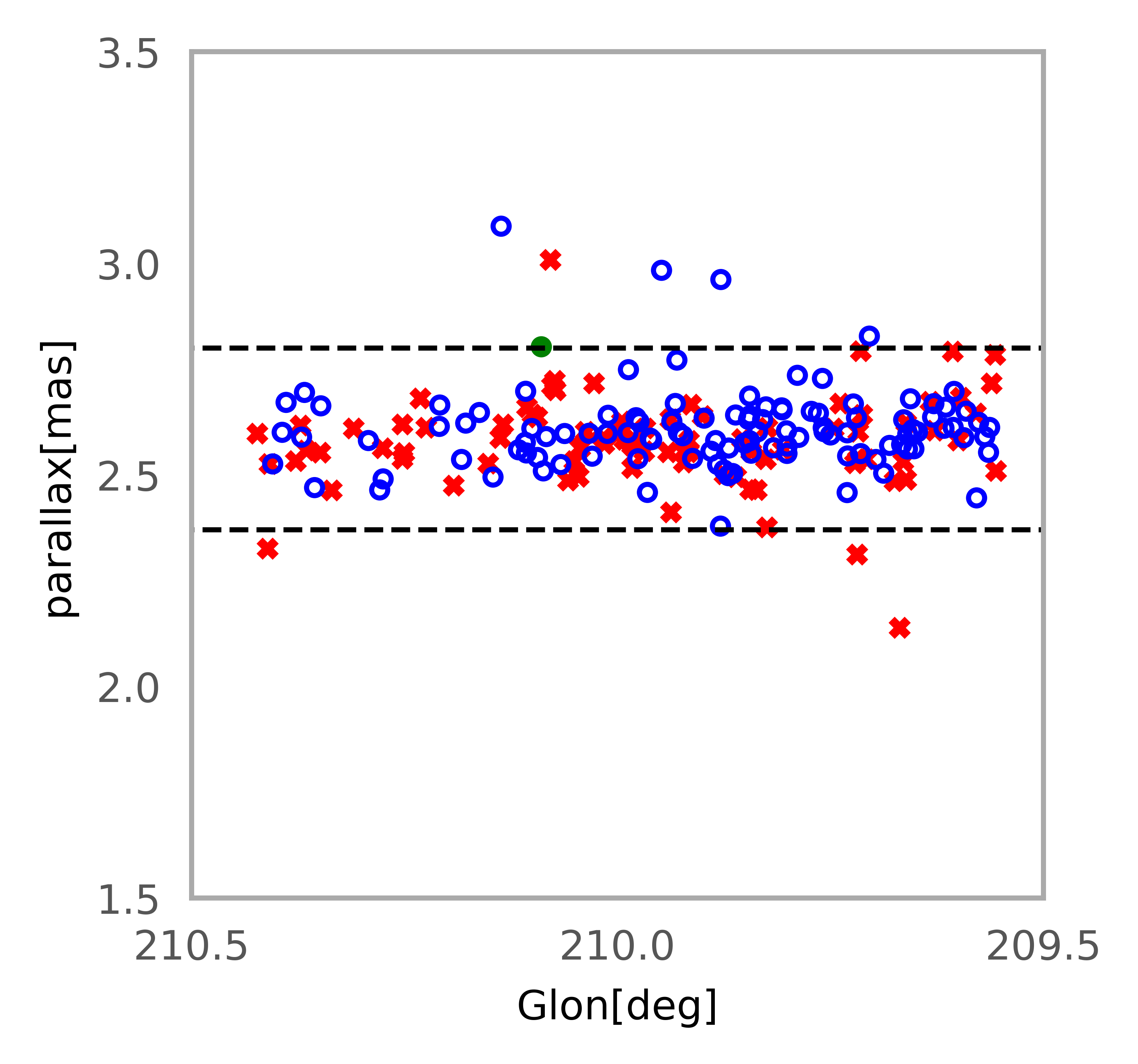}\label{l1641np}}
	\subfigure[L1641C]{\includegraphics[scale=0.6]{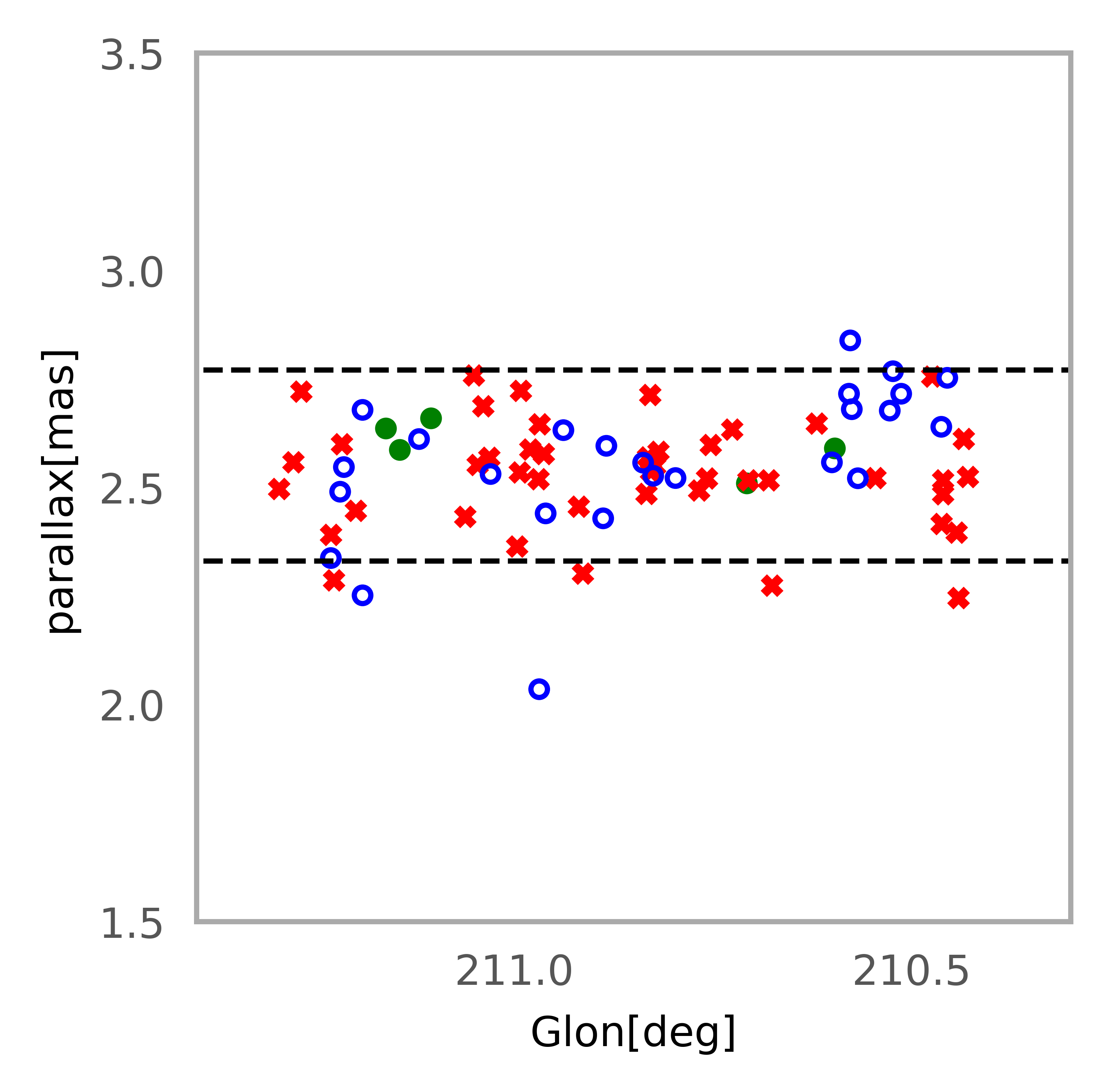}\label{l1641cp}}
	\subfigure[L1641S1]{\includegraphics[scale=0.6]{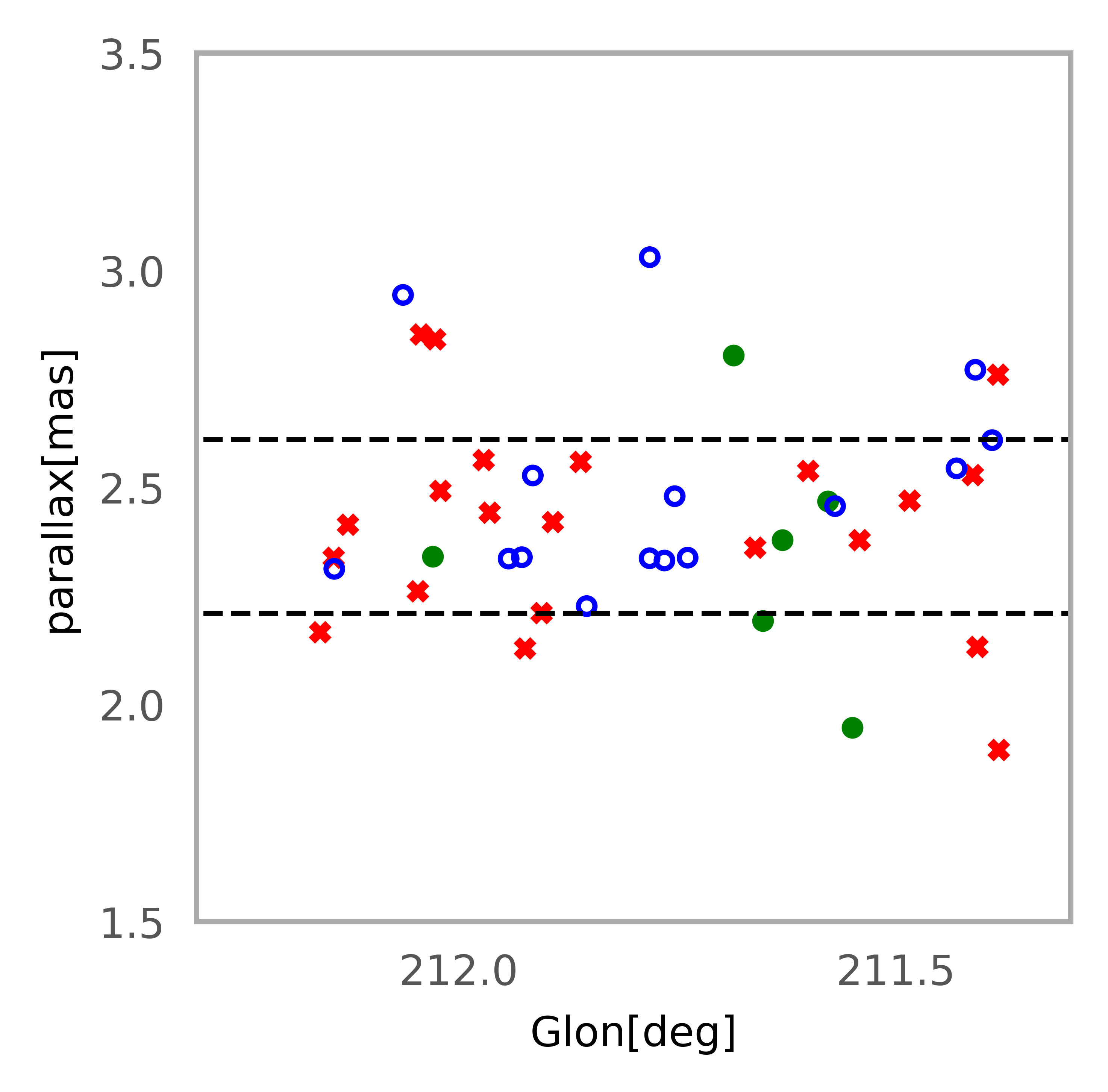}\label{l1641s1p}}
	\subfigure[L1641S2]{\includegraphics[scale=0.6]{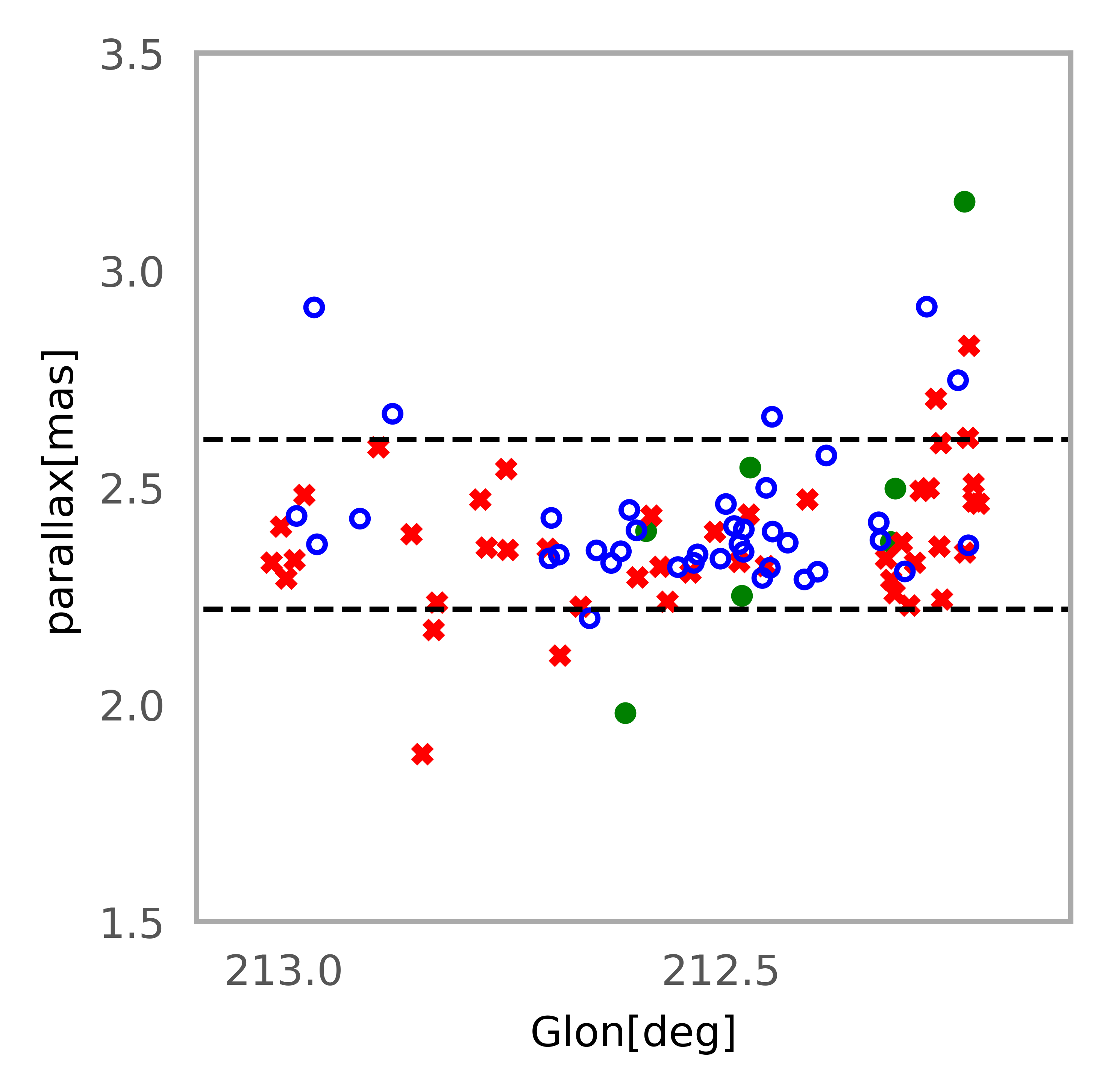}\label{l1641s2p}}
	\caption{Parallax distributions and criteria applied to the YSOs in the Orion~A, shown for (a) the OMC, (b) L1641N, (c) L1641C, (d) L1641S1, and (e) L1641S2. 
		The green solid circles represent the Class I YSOs, the red crosses represent the Class II YSOs, and the blue open circles represent the Class III YSOs.}
	\label{fig:parallax1}
\end{figure}

\begin{figure}[htbp]
	\centering
	\subfigure[L1630N]{\includegraphics[scale=0.6]{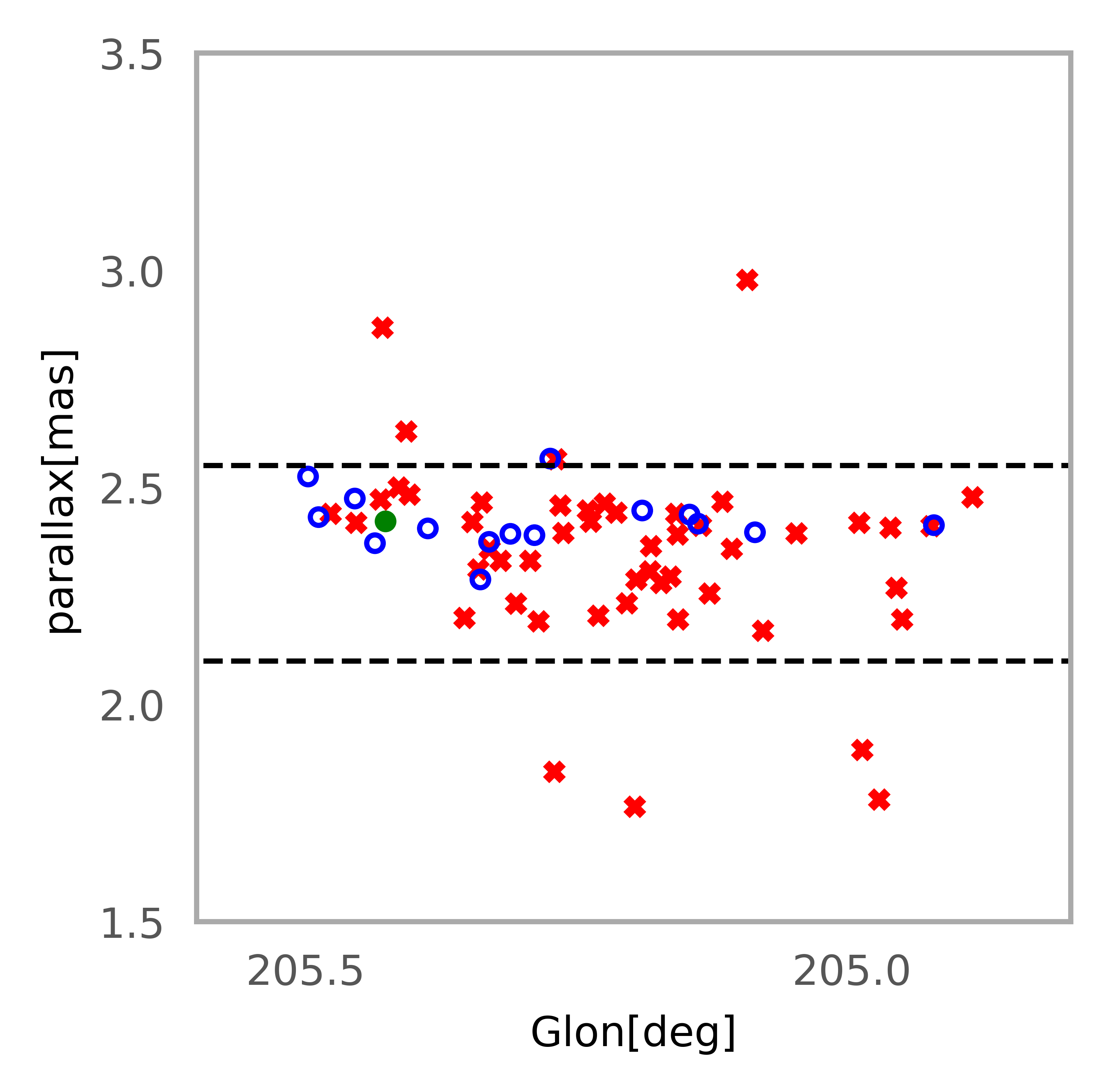}\label{l1630np}}
	\subfigure[NGC 2024]{\includegraphics[scale=0.6]{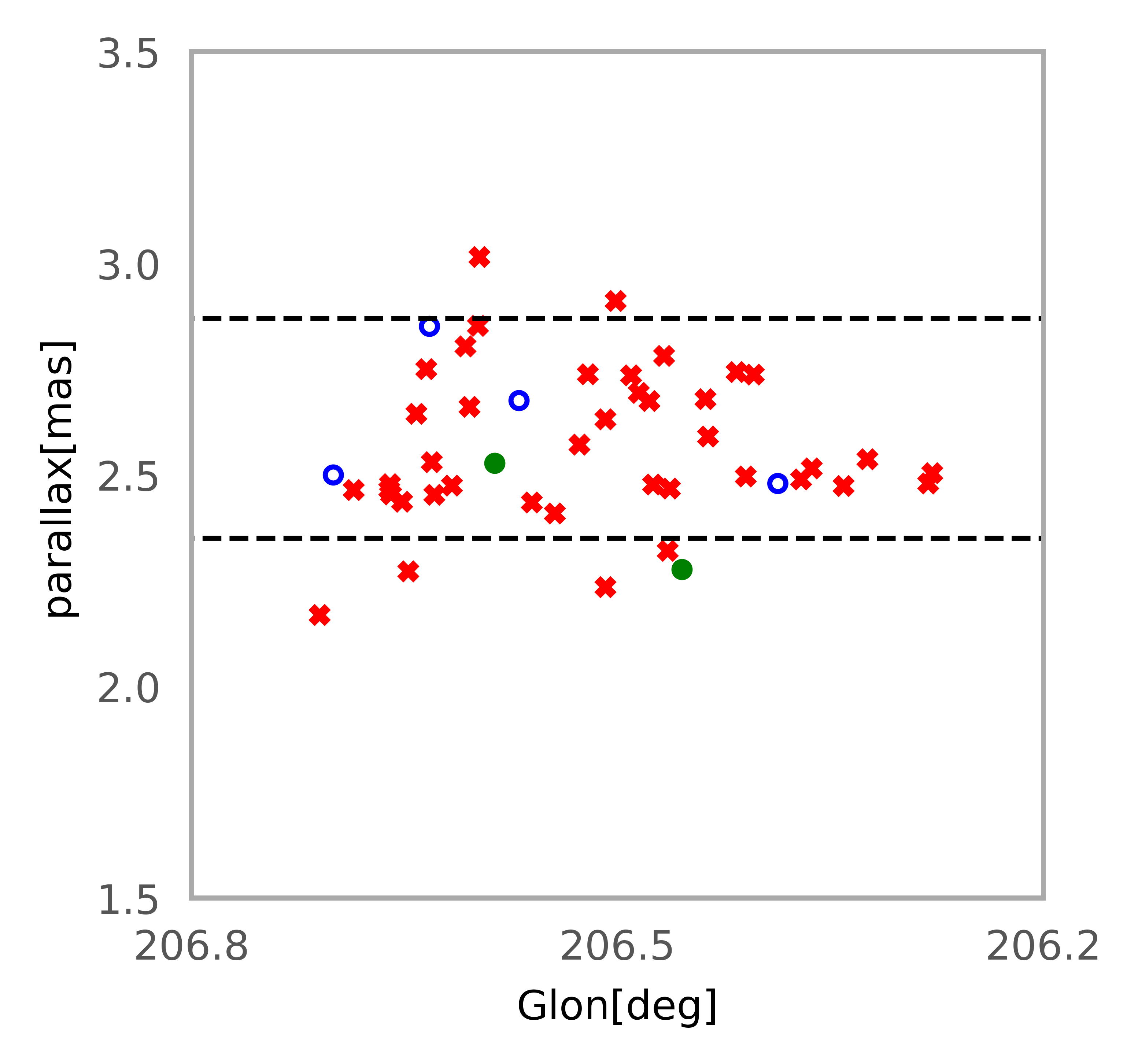}\label{ngc2024p}}
	\caption{Parallax distributions and criteria applied to the YSOs in Orion~B, shown for (a) L1630N and (b) NGC~2024. The other features are the same as in Fig. \ref{fig:parallax1}.}
	\label{fig:parallax2}
\end{figure}

\begin{figure}[htbp]
	\centering
	\subfigure[IC 348]{\includegraphics[scale=0.6]{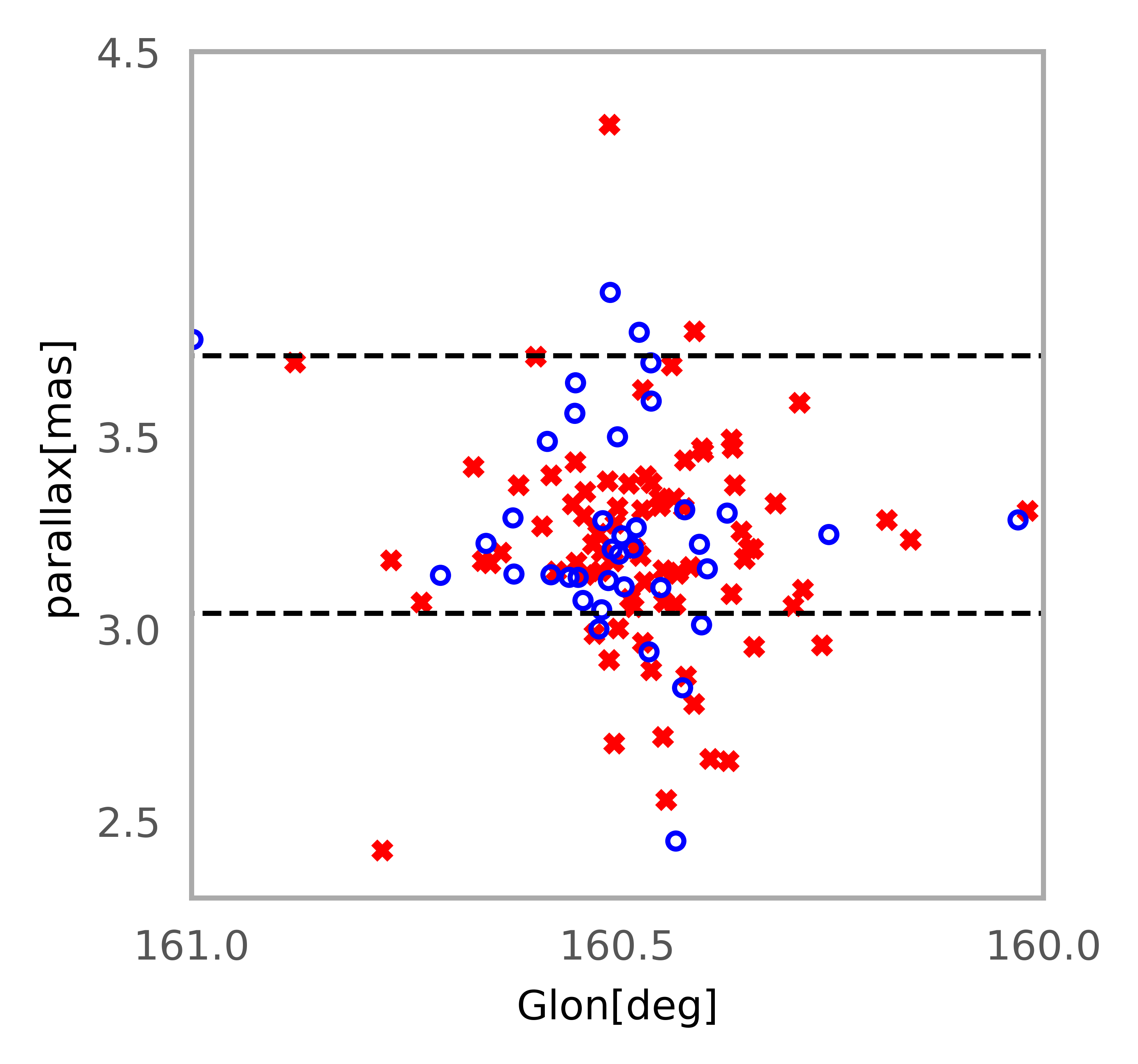}\label{ic348p}}
	\subfigure[NGC 1333]{\includegraphics[scale=0.6]{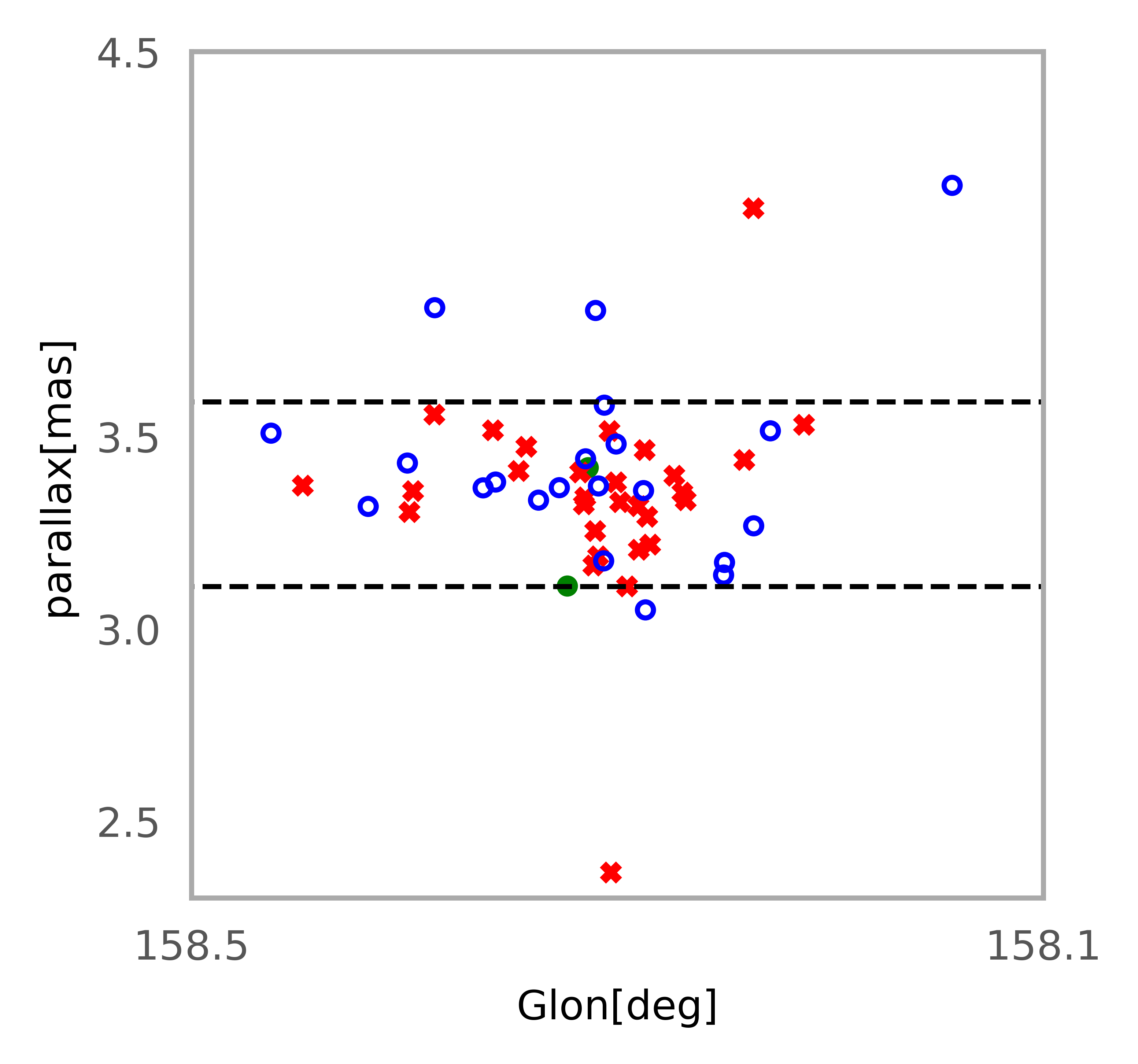}\label{ngc1333p}}
	\caption{Parallax distributions and criteria applied to the YSOs in Perseus, shown for (a) IC~348 and (b) NGC~1333. The other features are the same as in Fig. \ref{fig:parallax1}.}
	\label{fig:parallax3}
\end{figure}

\begin{figure}[htbp]
	\centering
	\subfigure[Heiles' Cloud 2]{\includegraphics[scale=0.6]{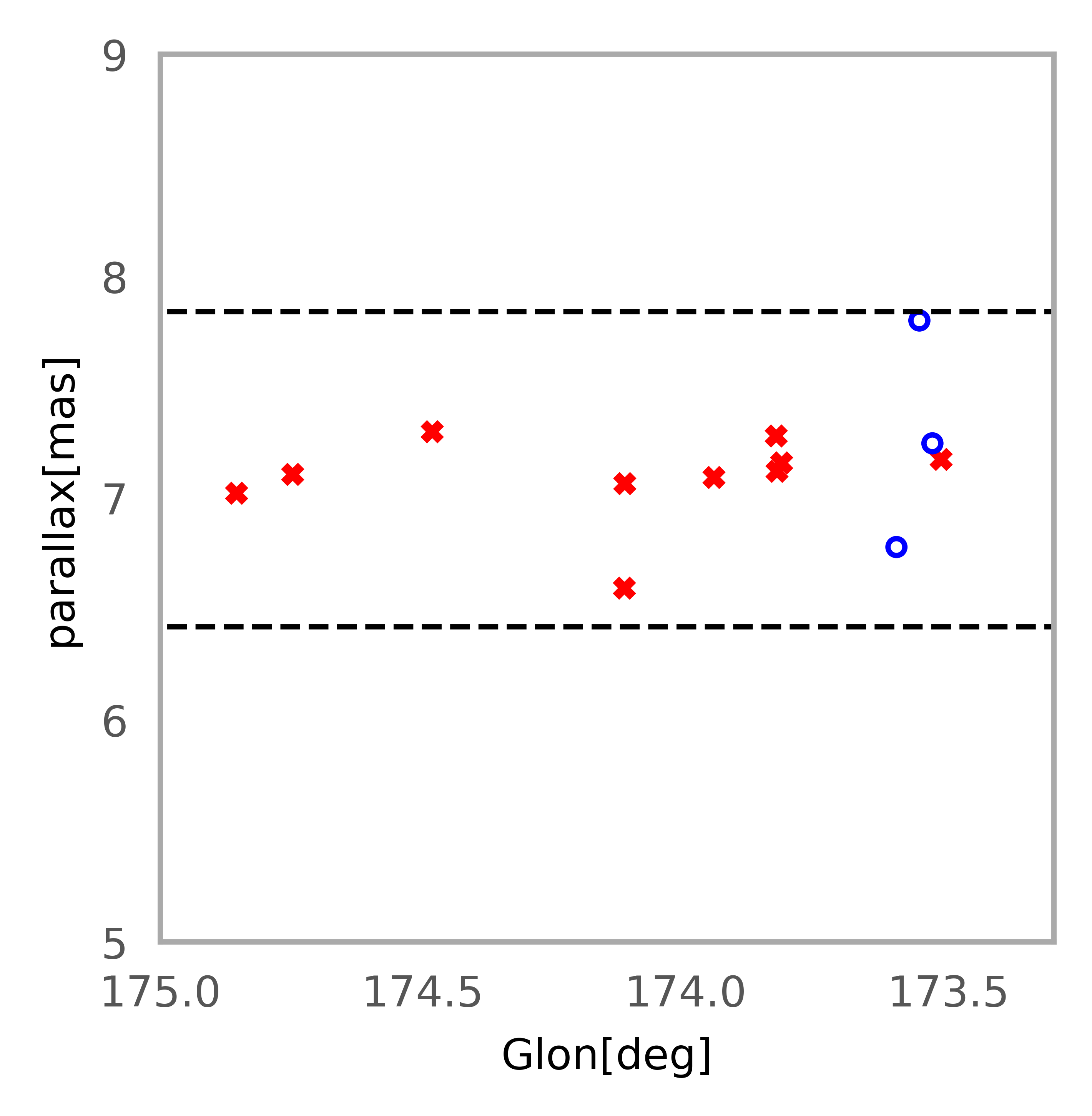}\label{hc2p}}
	\subfigure[L1536]{\includegraphics[scale=0.6]{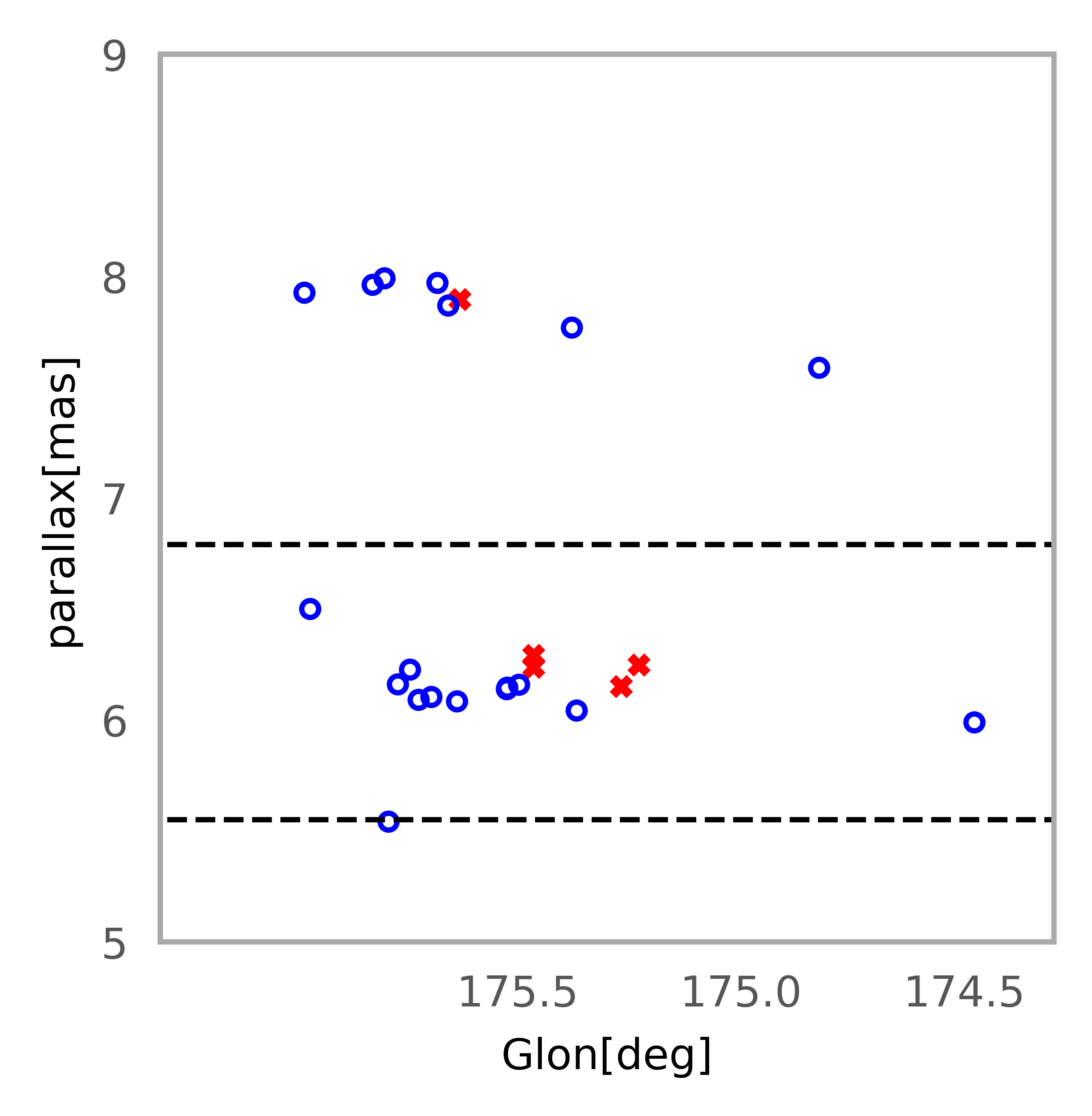}\label{l1536p}}
	\subfigure[B18]{\includegraphics[scale=0.6]{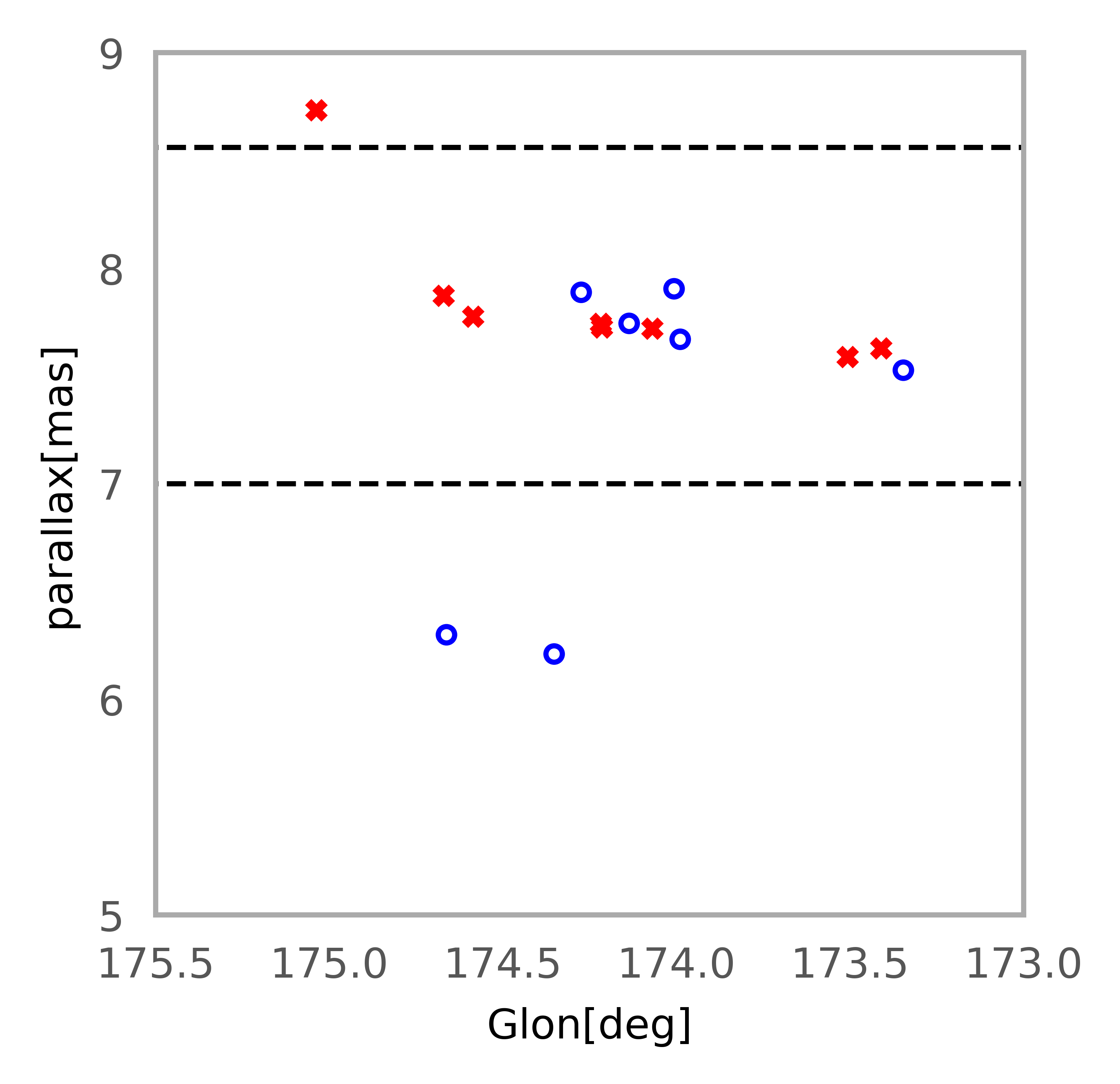}\label{b18p}}
	\subfigure[L1495]{\includegraphics[scale=0.6]{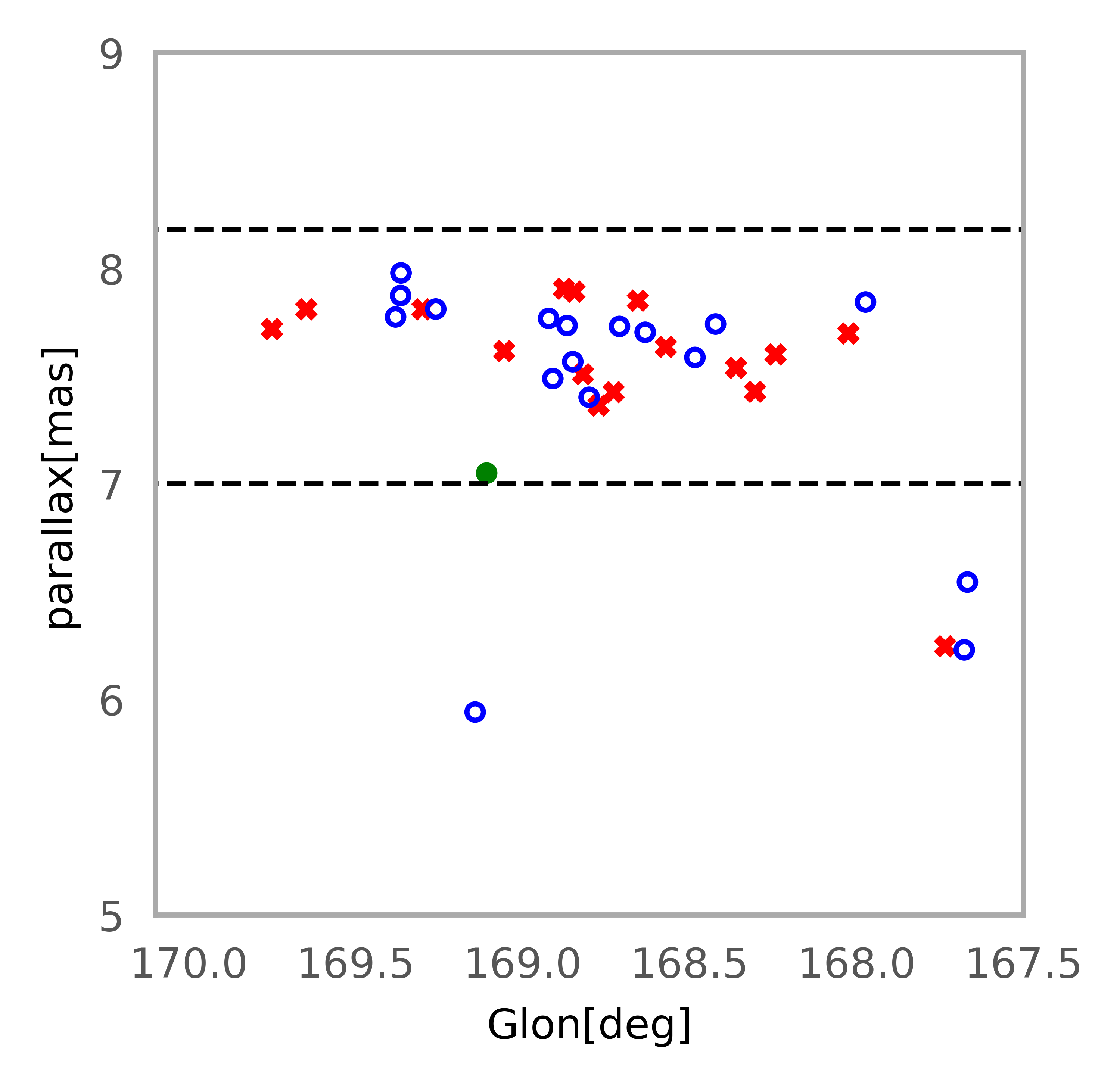}\label{l1495p}}
	\subfigure[B213]{\includegraphics[scale=0.6]{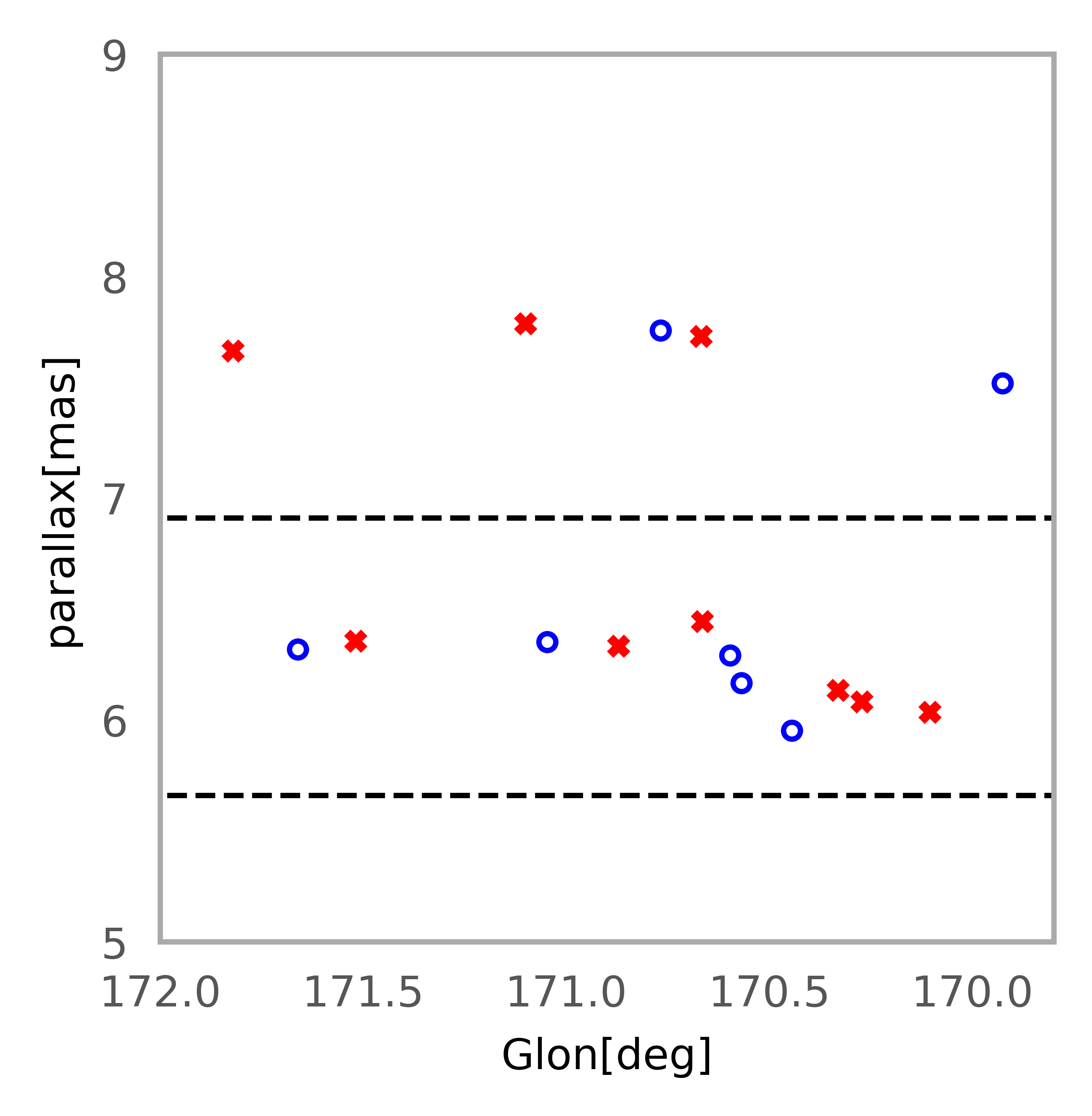}\label{b213p}}
	\caption{Parallax distributions and criteria applied to the YSOs in Taurus, shown for (a) Heiles' Cloud 2, (b) L1536, (c) B18, (d) L1495, and (e) B213. The other features are the same as in Fig. \ref{fig:parallax1}.}
	\label{fig:parallax4}
\end{figure}

\begin{figure}[htbp]
	\centering
	\subfigure[B30]{\includegraphics[scale=0.6]{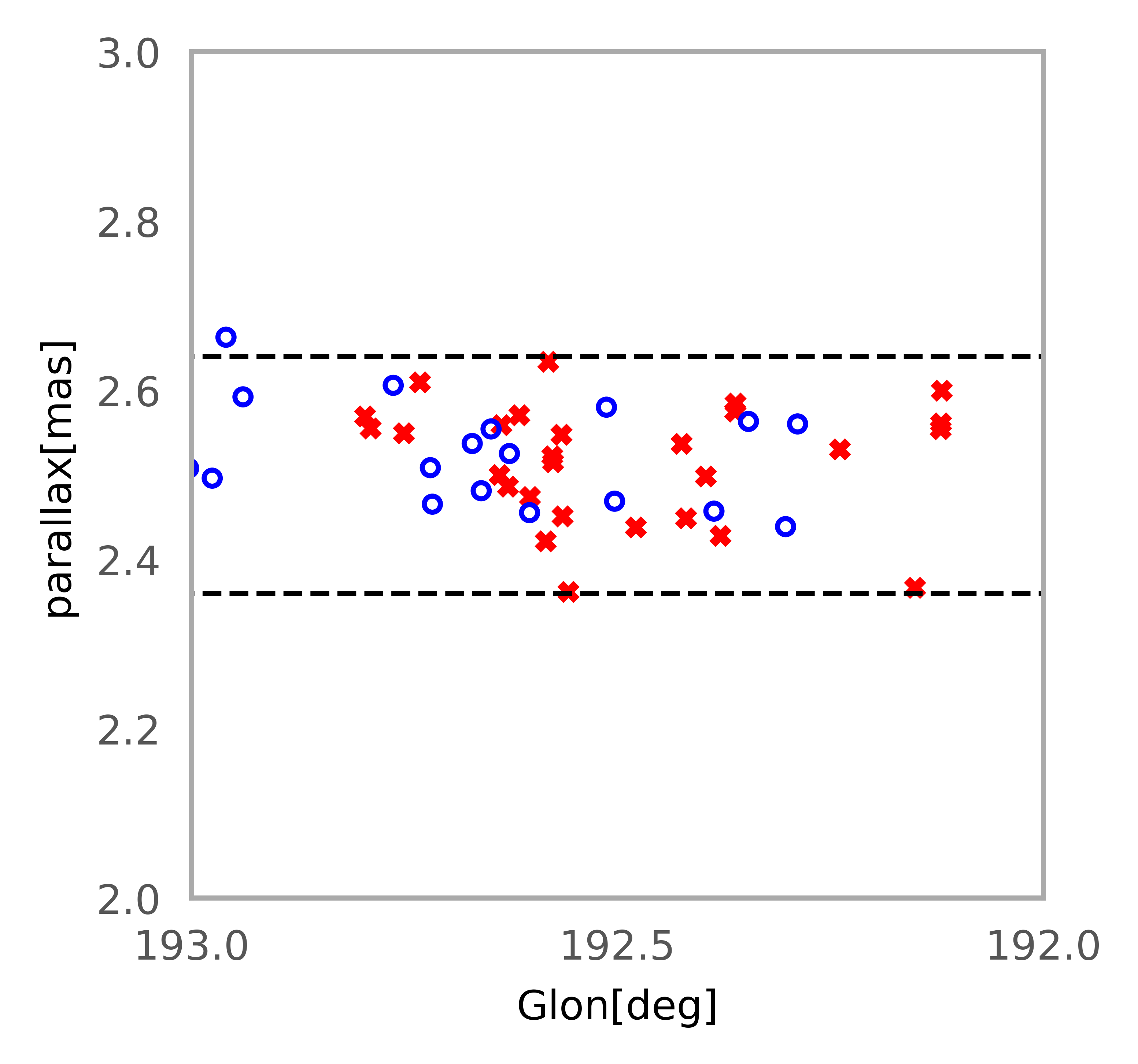}\label{b30p}}
	\caption{Parallax distributions and criteria applied to the YSOs in the $\lambda$~Orionis. The other features are the same as in Fig. \ref{fig:parallax1}.}
	\label{fig:parallax5}
\end{figure}

Figures \ref{fig:parallax1}--\ref{fig:parallax5} show the parallax distributions and criteria for the YSO samples in each subregion after performing our initial selection steps. After crossmatching with the source catalog, and applying the parallax and velocity criteria, we calculated the average distances of the YSO samples within each subregion. The upper and lower limit of the YSO sample distances were then determined to be $\pm10\%$ of the average distance. The parallax criteria for all subregions are listed in Table \ref{tab:parallax limit}.

\bibliography{sample631}{}
\end{document}